\documentclass[groupedaddress,
 reprint,
 amsmath,amssymb,
 nofootinbib,
 superscriptaddress,
 aps,
 prd,
]{revtex4-1}

\usepackage{graphicx}
\usepackage{dcolumn}
\usepackage{bm}
\usepackage{hyperref}
\usepackage[utf8]{inputenc}
\usepackage[capitalise]{cleveref}
\usepackage{aas_macros}
\usepackage[normalem]{ulem}
\usepackage{color}

\usepackage[makeroom]{cancel}
\usepackage{printlen}

\usepackage{mathtools}
\usepackage{physics}

\begin{document}

\title{Gravitational Collapse in the Post-Inflationary Universe}

\author{Benedikt Eggemeier}
\email{benedikt.eggemeier@phys.uni-goettingen.de}
\affiliation{
 Institut f\"ur Astrophysik, Georg-August-Universit\"at G\"ottingen, D-37077 G\"ottingen, Germany
}

\author{Bodo Schwabe}
\email{bschwabe@unizar.es}
\affiliation{
 Institut f\"ur Astrophysik, Georg-August-Universit\"at G\"ottingen, D-37077 G\"ottingen, Germany
}
\affiliation{CAPA \& Departamento de F\'isica Te\'orica, Universidad de Zaragoza, 50009 Zaragoza}

\author{Jens C. Niemeyer}
\email{jens.niemeyer@phys.uni-goettingen.de}
\affiliation{
 Institut f\"ur Astrophysik, Georg-August-Universit\"at G\"ottingen, D-37077 G\"ottingen, Germany
}

\author{Richard Easther}
\email{r.easther@auckland.ac.nz}
\affiliation{Department of Physics, University of Auckland, Private Bag 92019, Auckland, New Zealand}

\date{\today}

\begin{abstract}
The Universe may pass through an effectively matter-dominated epoch between inflation and Big Bang Nucleosynthesis during which gravitationally bound structures can form on subhorizon scales. In particular, the inflaton field can collapse into inflaton halos, forming ``large scale'' structure in the very early universe. We combine N-body simulations with high-resolution zoom-in regions in which the non-relativistic Schrödinger-Poisson equations are used to resolve the detailed, wave-like 
structure of inflaton halos. 
Solitonic cores form inside them, matching structure formation simulations with axion-like particles in the late-time universe. We denote these objects \textit{inflaton stars}, by analogy with boson stars.
Based on a semi-analytic formalism we compute their overall mass distribution which shows that some regions will reach overdensities of $10^{15}$ if the early matter-dominated epoch lasts for 20 $e$-folds. 
The radii of the most massive inflaton stars can shrink below the Schwarzschild radius, suggesting that they could form primordial black holes prior to thermalization.
\end{abstract}

\maketitle

\section{Introduction}
\label{sec:intro} 

Cosmological  inflation~\cite{Starobinsky1980,Guth1981,Linde1982,Linde1983} is a period of accelerated expansion in the early Universe. In simple scenarios this phase ends with the scalar inflaton field oscillating about the (possibly local) minimum of its potential. A fully viable inflationary model must  contain a mechanism that converts the energy-density represented by the resulting inflaton condensate into Standard Model particles, thermalizing the universe at temperatures high enough  to allow neutrino freeze-out and nucleosynthesis. 

Many mechanisms could accomplish this task, given that in simple scenarios inflation typically ends with densities near the Grand Unification scale but thermalization is not absolutely required until interaction energies approach MeV scales.  In practice, thermalization is usually assumed to occur at the TeV scale or above, given that baryogenesis and dark matter production must occur prior to nucleosynthesis, and both processes presumably involve physics beyond the Standard Model. However, even in this case  characteristic  energies could vary by a factor of $10^{12}$ between the end of inflation and the onset of thermalization. 

A number of mechanisms can fragment the initially near-homogeneous, oscillating inflaton condensate. Quanta of fields coupled to the inflaton or the inflaton itself can be produced by resonance~\cite{Traschen:1990sw,Shtanov:1994ce,Kofman1997}, leading to quasi-exponential growth in the occupation numbers of specific ranges of momentum modes, rendering the early universe highly inhomogeneous. These ``preheating'' mechanisms typically produce a non-thermal initial distribution which would then be thermalized by their mutual interactions~\cite{Lozanov:2016hid}.  In addition, resonance can lead to long-lived collective excitations such as oscillons~\cite{Gleiser:1993pt,Copeland:1995fq,Amin2010,Amin2012,Lozanov:2017hjm} which are stable on scales much longer than the post-inflationary Hubble time.

In this work, we focus on scenarios in which the primary inflaton  interactions are gravitational.  In these cases, inflation is followed by an extended period of early matter domination (EMD); integrating out the rapid inflaton oscillations leads to an  effective description of density perturbations on a non-relativistic, matter-dominated background. Small perturbations in the inflaton condensate become gravitationally unstable inside the Hubble horizon and collapse~\cite{Easther2010,Jedamzik:2010hq}, eventually forming bound structures~\cite{Musoke2019,Niemeyer2019,Eggemeier2020}.  

It is possible that this era supports a lengthy period during which the local gravitational dynamics are nonlinear, leading to large overdensities~\cite{Niemeyer2019,Eggemeier2020}. These structures are necessarily evanescent, since thermalization must convert all remnant inflaton material into Standard Model particles and dark matter. The details of this phase are convolved with the predicted values of observables associated with primordial perturbations~\cite{Liddle2003,Adshead2008,Adshead2011}. Furthermore, this complex nonlinear phase could conceivably source stochastic gravitational waves via merging halos, primordial black hole formation, as well as inhomogeneous reheating prior to nucleosynthesis.

Given that this epoch involves fully nonlinear gravitational clustering three-dimensional simulations are required to model its detailed evolution. The non-relativistic description of a generic self-gravitating scalar field  is provided by the Schrödinger-Poisson equations. Gravitational clustering in the post-inflationary epoch was first demonstrated  in Ref.~\cite{Musoke2019} by numerically solving the Schrödinger-Poisson equations on a comoving background. As  with axion-like or fuzzy dark matter (FDM)~\cite{Hu2000} structure formation simulations, the need to spatially resolve the de Broglie wavelength limits  the achievable  simulation volume~\cite{Schwabe2016, Niemeyer2019_review}. On scales far greater than the de Broglie length, the coarse grained behavior of scalar fields converges to that of collisionless particles~\cite{Widrow1993,Uhlemann2014}, allowing the use of standard N-body methods if only large-scale dynamics are of interest. 
The large-scale correspondence between the coarse grained Schrödinger-Poisson and the Vlasov-Poisson equations was exploited to analyse the gravitational fragmentation of the inflaton field after inflation in Ref.~\cite{Eggemeier2020}.
Large N-body simulations showed that the inflaton condensate collapses into gravitationally bound inflaton halos, reaching masses of up to $20\,\mathrm{kg}$ with radii of the order of $10^{-20}\,\mathrm{m}$. The resulting inflaton halo mass function (IHMF) is consistent with results from Press-Schechter theory~\cite{Niemeyer2019}. 

N-body methods are computationally efficient 
but do not capture phenomena arising from wave interference patterns, whose characteristic scale is set by the de Broglie wavelength. These include the formation of gravitationally bound solitonic objects equivalent to non-relativistic Bose stars. These have received  attention in the context of FDM structure formation~\cite{Schive:2014dra,Schive:2014hza,Schwabe2016,Veltmaat:2018dfz,Mocz2017,Mocz2018} and QCD axion minihalos~\cite{Tkachev1986,Tkachev1991,Levkov2018,Eggemeier2019PRD} where they are known as \emph{solitonic halo cores} and \emph{(dilute) axion stars}, respectively. The similarity of initial conditions on sub-horizon scales -- a homogeneous, cold field with small density perturbations --  suggests that the Schrödinger-Poisson dynamics of the inflaton field during EMD  give rise to equivalent bound structures, referred to as \emph{inflaton stars} in Ref.~\cite{Niemeyer2019}. 

Confirming this hypothesis requires simulations using the Schrödinger-Poisson equations on small scales with appropriate initial conditions for mass density and momentum in a  cosmological box large enough to follow the evolution of structures akin to the present-day cosmic web. The analogous problem  in the context of FDM structure formation was solved by using a hybrid approach with adaptive-mesh refinement (AMR) techniques~\cite{Veltmaat:2018dfz}. Using an N-body particle representation of the scalar field on coarse grid levels and only solving the Schrödinger-Poisson equations in isolated, highly refined regions surrounding pre-selected halos made it possible to observe the formation and growth of solitonic cores from cosmological initial conditions. 

We use the same hybrid methods to extend our previous N-body simulations of the early matter-dominated post-inflationary era~\cite{Eggemeier2020} to much smaller length scales. This confirms the formation of inflaton stars at the center of inflaton halos during EMD. 
Using the IHMF from our previous N-body simulations and the knowledge that these solitons are in virial equilibrium with their host halos we determine the inflaton star mass function (ISMF) and find that it can be well approximated by a modified Press-Schechter approach~\cite{Press1974}. 
With overdensities of up to $10^{15}$ the existence of inflaton stars might lead to new possible observational hints of the early matter-dominated epoch. 
If the universe thermalizes at the TeV scale, typical overdensities of $10^{32}$ will be reached and the most massive inflaton stars can collapse into primordial black holes (PBHs).

The structure of this paper is as follows. In \cref{sec:methods} we review the formation of gravitational bound structures after inflation and describe our numerical methods and  simulation setup. We present our results in \cref{sec:results}. The mass distribution of inflaton stars and the possible formation of black holes is discussed in \cref{sec:inflatonstar_MF}. We conclude in \cref{sec:conclusions}. 

\section{Simulation Setup and Numerical Methods} 
\label{sec:methods}

In single-field models the accelerated expansion of the Universe is driven by the scalar inflaton field $\varphi$ with potential $V(\varphi)$. In the homogeneous limit of a flat Friedmann-Lemaitre-Robertson-Walker (FLRW) space-time, the inflaton obeys
the Klein-Gordon equation
\begin{align}
    \Ddot{\varphi} + 3H\dot\varphi + \frac{\mathrm{d}V}{\mathrm{d}\varphi} = 0\,,
    \label{eq:Klein-Gordon}
\end{align}
while the Friedmann equation
\begin{align}
    H^2 = \frac{1}{3M_\mathrm{Pl}^2}\left(\frac{1}{2}\dot\varphi^2 + V(\varphi)\right)
    \label{eq:friedmann}
\end{align}
describes the expansion of space. Here, $M_\mathrm{Pl} = (8\pi G)^{-1/2}$ is the reduced Planck mass and $H= \dot a/a$ the Hubble parameter with scale factor $a$. 

After slow-roll inflation, the homogeneous inflaton condensate oscillates around its potential minimum. While the full potential must be shallower than quadratic at large field values $\varphi > M_\mathrm{Pl}$ for compatibility with observations~\cite{Planck2018_inflation}, most inflationary potentials can be described by $V(\varphi)\sim\varphi^2$ around potential minima, or
\begin{align}
    V(\varphi) = \frac{1}{2}m^2\varphi^2\,.
    \label{eq:potential}
\end{align}
The power spectrum of density perturbations for this potential decays as $k^{-5}$ on subhorizon scales~\cite{Easther2010,Eggemeier2020}.

Assuming that the full potential does not support resonance and 
solving \cref{eq:Klein-Gordon} we find that in the quadratic limit the inflaton field evolves as $\varphi\sim \sin(mt)/t$ in the post-inflationary epoch. Furthermore, the scale factor behaves as $a(t)\sim t^{2/3}$ and the Hubble parameter decreases as $H\sim a^{-3/2}$, similar to the expansion in a matter-dominated universe~\cite{Albrecht:1982mp}.

If the inflaton coupling to other fields is small, this matter-dominated era can last for multiple $e$-folds. For example with a reheating temperature of $T_\mathrm{rh}\simeq 10^7\,\mathrm{GeV}$, the universe grows by 24 $e$-folds~\cite{Eggemeier2020}. During this period, sub-horizon density perturbations grow linearly with the scale factor and eventually collapse into gravitationally bound structures~\cite{Easther2010, Jedamzik:2010hq}.

At the end of slow-roll inflation, $\varphi_\mathrm{end}\sim M_\mathrm{Pl}$ and thus $H_\mathrm{end}\sim m$, so that a few $e$-folds after the end of inflation we have $m\gg H$,
allowing the non-relativistic approximation of the Klein-Gordon equation on subhorizon scales.
Additionally, large occupation numbers allow a classical treatment of the inflaton field, which is often referred to as a condensate. In the absence of self-interactions or interactions with other fields, the inflaton condensate evolves solely under the influence of its own self-gravity. 
As demonstrated in Ref.~\cite{Musoke2019}, these conditions justify the use of the WKB approximation in that
the Klein-Gordon-Einstein equation approaches its non-relativistic limit, the (comoving) Schrödinger-Poisson equations~\cite{Ruffini1969,Nambu1990QuantumPerturbations}:
\begin{align}
     i\hbar \partial_t\psi &= -\frac{\hbar^2}{2ma^2}\nabla^2\psi +  m \psi V_N\,, \label{eq:Schroedinger_eq}\\
     \nabla^2V_N &= \frac{4\pi G}{a}\left(\rho - \bar\rho\right)\,.
     \label{eq:poisson_eq}
\end{align}
Here, $\rho = \abs{\psi}^2$ is the density of the scalar field with mean density $\bar\rho$. 

Numerical simulations of the Schrödinger-Poisson equations~\cite{Musoke2019} demonstrated the gravitational fragmentation and clustering of the inflaton field, but limited spatial resolution and box size prevented the computation of halo statistics or the detailed simulation of solitonic core formation\footnote{In principle one could evolve the inflaton field via the Einstein-Klein-Gordon equations but it is computationally costly to do this over  many $e$-folds and there is also no reason to use the full formalism, given that the non-relativistic approximation is valid.}. Their properties were estimated in Ref.~\cite{Niemeyer2019} using a Press-Schechter analysis and the core-halo mass relation measured in FDM simulations~\cite{Schive:2014hza}, predicting characteristic masses of $10^{-3}\,\mathrm{kg}$ for inflaton halos and $10^{-6}\,\mathrm{kg}$ for their solitonic cores (inflaton stars), with typical overdensities reaching $\mathcal{O}(10^{6})$.

Large N-body simulations in the Vlasov regime of gravitational clustering  were reported in Ref.~\cite{Eggemeier2020}. They confirmed the formation of inflaton halos with halo mass functions consistent with Press-Schechter models with a sharp-$k$ filter. Moreover, the averaged halo density profiles were shown to agree well with NFW-fits, providing further evidence for the structural equivalence of nonlinear gravitational clustering in EMD and late-time cosmological structure formation in the collisionless kinetic regime. 

We extend these N-body simulations by finely resolving pre-selected isolated inflaton halos to below the de Broglie wavelength $\lambda_\mathrm{dB} = 2\pi\hbar/(mv_\mathrm{vir})$ that is set by their virial velocity $v_\mathrm{vir}$. Implementing the hybrid method introduced in Ref.~\cite{Veltmaat:2018dfz} within \textsc{AxioNyx}~\cite{Schwabe2020}
enables us to solve the Schrödinger-Poisson equations on adaptively refined regular grids while using the standard N-body solver for most of the simulation volume. This reduces the computation time significantly while resolving the wave-like properties of the inflaton field in regions of interest, which cover only $1.5\times 10^{-3}\%$ of the full simulation domain.

\subsection{Classical wave approximation}

\begin{figure*}
    \centering
    \includegraphics[width=\textwidth]{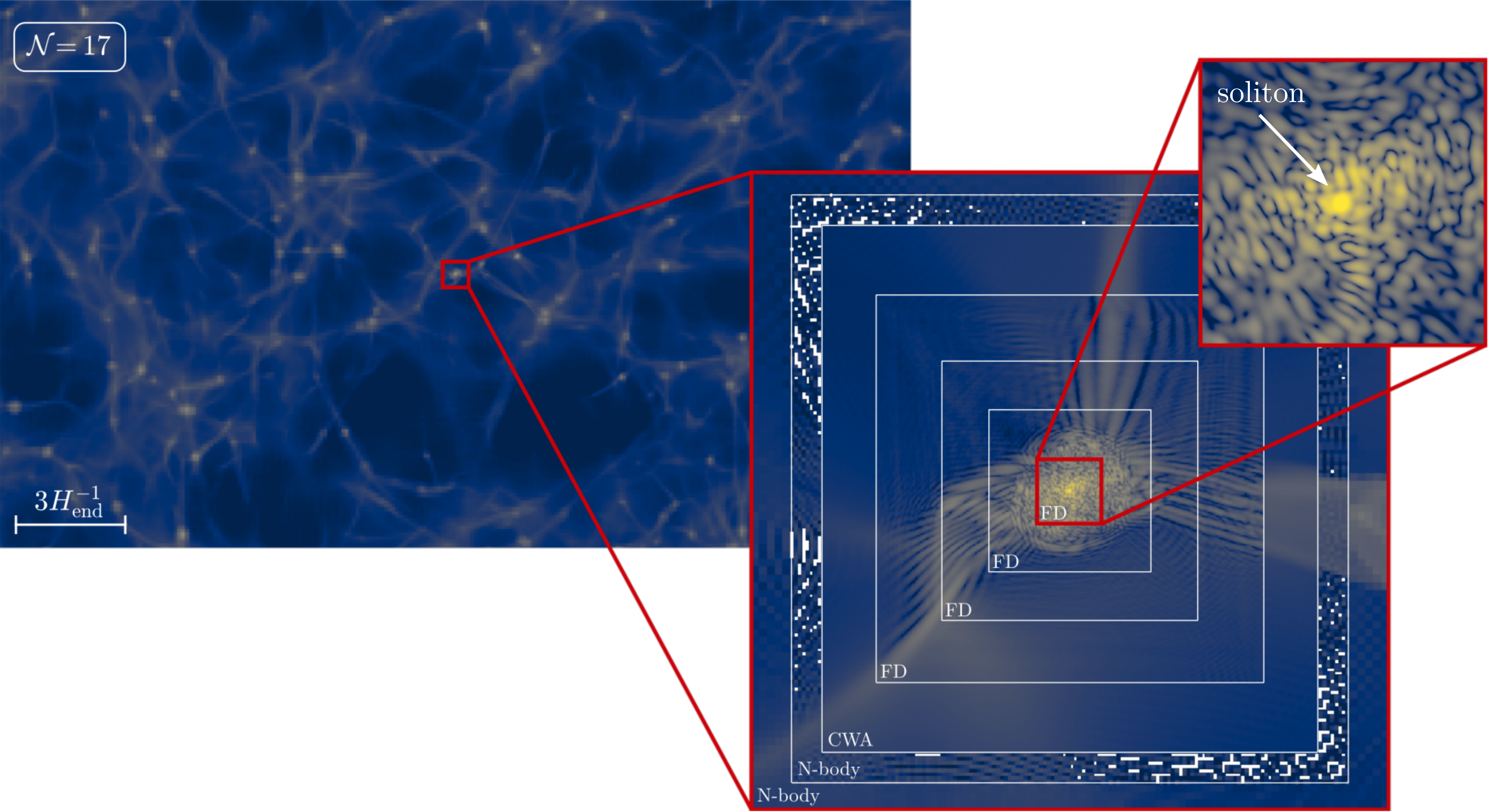}
    \caption{Scaled sequence of the inflaton density $\mathcal{N}=17$ $e$-folds after the end of inflation. The rectangular region shows the projected inflaton density of $15\%$ of the full simulation box centered on a selected halo. A slice through the maximum density of the halo illustrates the setup of several refinement levels in our simulation. Note that the white space in the last N-body level correspond to cells without any particles. Compared to the root grid the spatial resolution of the finest finite-difference (FD) level, which is displayed in an enlargement in the upper right panel, is increased by a factor of $2^8=256$. One can clearly recognize the interference patterns in the filaments, the granular structure inside the collapsed inflaton halo and the solitonic core in its center. }
    \label{fig:dens_zoom}
\end{figure*}

The hybrid method relies on the reconstruction of the wave function from particle information at the N-body/Schrödinger-Poisson boundary.  We employ the classical wave approximation (CWA) described in Ref.~\cite{Veltmaat:2018dfz} for FDM structure formation to achieve this transition. In this hybrid scheme a classical wave function (CWF) is created from  information provided by N-body particles. Apart from their masses $m_i$, positions $\mathbf{x}_i$ and velocities $\mathbf{v}_i$, these particles also need to carry information about the wave function's complex phase. 
Applying periodic boundary conditions to the Schrödinger-Poisson equations, the initial phases $S_i$ of the N-body particles in the simulation volume can be obtained from the initial particle velocity field $\mathbf{v} = a^{-1}\nabla S$ (in physical units) by solving
\begin{align}
    \nabla\cdot \mathbf{v} = a^{-1}\nabla^2 S\,,
\end{align}
with \textsc{AxioNyx}'s Poisson solver.  
After initialization, the particle phases are evolved in each time step
according to the Hamilton-Jacobi equation~\cite{Veltmaat:2018dfz}
\begin{align}
    \frac{\mathrm{d} S_i}{\mathrm{d}t} = \frac{1}{2}\mathbf{v}_i^2 - V_N(\mathbf{x}_i)\,.
\end{align}

Before a selected halo starts to collapse, the CWF is constructed 
in each subsequent time step at the N-body/Schrödinger-Poisson boundary 
encompassing the halo under the assumption
that interference effects can be neglected.
This is justified in low-density regions with linear dynamics, where the Schrödinger-Vlasov correspondence is valid and wave-like effects are suppressed, but does not hold in the multi-streaming regime of a collapsing halo or filament. It is therefore important to construct the CWF before the collapse of non-linear structures and in a large enough volume in order to avoid boundary effects in regions of interest (cf. \cref{fig:dens_zoom}).
Before each time step on the finite-difference level where the Schrödinger-Poisson equations are solved, the amplitude and the phase of the CWF are constructed at the boundaries.
Its amplitude is given by~\cite{Veltmaat:2018dfz}
\begin{align}
    \abs{\psi(\mathbf{x})} = \left(\sum_i W(\mathbf{x} - \mathbf{x}_i)\right)^{1/2}\,,
    \label{eq:cwa_amplitude}
\end{align}
where the particle masses are smoothed by the mass conserving interpolation kernel
\begin{align}
    W(\mathbf{x} - \mathbf{x}_i) = m_i \frac{3}{\pi\xi^3}\left(1 - \frac{\abs{\mathbf{x} - \mathbf{x}_i}}{\xi}\right)
\end{align}
for $\abs{\mathbf{x} - \mathbf{x}_i} < \xi$ and 0 elsewhere. Thus, $\xi$ serves as a smoothing radius for the interpolation of the particles onto the grid.

The phase $S(\mathbf{x})$ is the argument of the superposition of particles weighted by their interpolation kernel and given by~\cite{Veltmaat:2018dfz}
\begin{align}
    S(\mathbf{x}) = \frac{\hbar}{m}\mathrm{arg}\left(\sum_i\sqrt{W(\mathbf{x} - \mathbf{x}_i)}e^{i[S_i + \mathbf{v}_i\cdot a(\mathbf{x} - \mathbf{x}_i)]m/\hbar}\right)\,.
    \label{eq:cwa_phase}
\end{align}

\subsection{Simulation Setup}

We use the same model parameters and unit system as in our previous N-body simulations of the matter-dominated ($\Omega_m = 1$) post-inflationary era~\cite{Eggemeier2020}. Explicitly, we assume the quadratic potential of \cref{eq:potential},  $\varphi_\mathrm{end} \approx M_\mathrm{Pl}$ and $H_\mathrm{end} \approx m/\sqrt{6}$ at the end of inflation, where $m=6.35\times 10^{-6}\,M_\mathrm{Pl}$ is the inflaton mass. The comoving length unit $l_u$ of our simulations is determined by the physical size of the horizon $\mathcal{N}=20$ $e$-folds after the end of inflation, i.e. $l_u = e^{20} H_\mathrm{end}^{-1} = 1.51\times 10^{-20}\,\mathrm{m}$. Setting the mass unit $m_u = 10^{-10}\,\mathrm{g}$ and choosing the time unit $t_u = 7.23\times 10^{-24}\,\mathrm{s}$, the gravitational constant is $G= 10^{-10}\,l_u^3/(m_u t_u^2)$. The Hubble parameter and the mean density $\mathcal{N}=20$ $e$-folds after inflation are $H_{20} = 6.49\,t_u^{-1}$ and $\bar\rho = 5.02\times 10^{10}\,m_u/l_u^3$, respectively. The comoving length of our simulation box is set to $L=50\,l_u$.

We make use of one further stratagem to ensure that the numerical simulations are tractable. The overall cosmological parameters are set relative to the choice $m=6.35\times 10^{-6}\,M_\mathrm{Pl}$, as noted above. However, we artificially increase $\lambda_\mathrm{dB}$ in the simulations 
by reducing $m$ by a factor of $80$,  so that $\hbar/m = 9.84\times  10^{-3}\,l_u^2/t_u$. This increases the lengths at which wave effects become apparent and reduces the number of nested levels we need to resolve, and allows a larger timestep within the simulation. Taken together these choices prevent the simulations from becoming computationally intractable.  This approach is justifiable given the scaling symmetry~\cite{Guzman2004} of the Schrödinger-Poisson equations which maps soliton quantities to rescaled inflaton masses. Hence, our results are qualitatively independent of the inflaton mass. 

\begin{figure*}
     \centering
     \includegraphics[width=\textwidth]{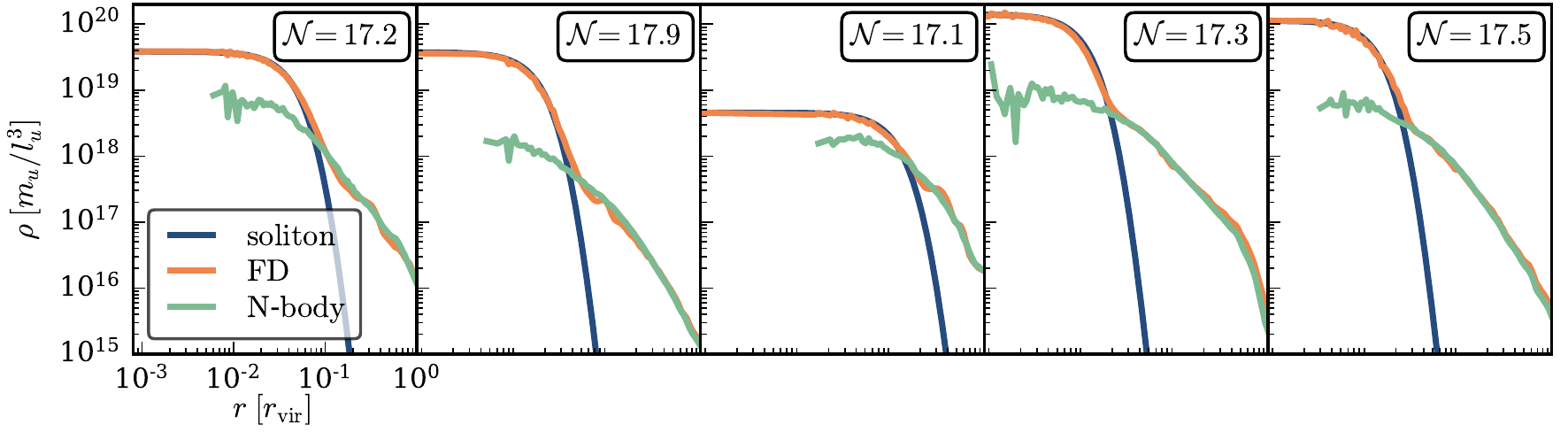}
     \caption{Radial density profiles of five simulated halos at different number of $e$-folds $\mathcal{N}$ after the end of inflation. 
     The orange curves represent the profiles from solving the Schrödinger-Poisson equations with the finite-difference method while the blue curves are given by the theoretical soliton density profiles from \cref{eq:soliton_profile}. For comparison, the halo profiles from a pure N-body simulation are shown in green. }
     \label{fig:densprofiles}
 \end{figure*}

We use the same matter power spectrum as in Ref.~\cite{Eggemeier2020} (see Fig.~2 therein) to create the initial conditions for our simulations starting $\mathcal{N}=14$ $e$-folds after the end of inflation. The power spectrum of density perturbations was computed numerically in Ref.~\cite{Easther2010} over a wide range of $k$.  Since the nonlinear evolution of the density perturbations is not sensitive to its precise form, it is sufficient to interpolate between the super- and subhorizon limits of the power spectrum~\cite{Niemeyer2019,Eggemeier2020}. 

The  $512^3$ unigrid N-body simulations described in Ref.~\cite{Eggemeier2020} are used to determine the Lagrange patches of five isolated halos that experience no major mergers during the entire simulation time. In five separate simulations, these regions are individually refined with two additional static refinement levels with a comoving side length of up to $7.5\,l_u$. The number of particles in these regions is increased accordingly. We employ \textsc{Music}~\cite{Hahn2011} to generate particle positions and velocities for these nested initial conditions.

Initially, the particles are  in the single-streaming regime and can be evolved  with the N-body solver until $\mathcal{N}\approx 16.5$ without further refinement. As the selected halos start to collapse, we add three additional refinement levels tracing the halos' central positions. The first is an additional N-body level, followed by a level with a cubic grid of comoving side length $1.5\,l_u$ on which the CWF is constructed with $\xi = 6\,\Delta x$, where $\Delta x$ denotes the level's cell width. The initial CWF is  interpolated onto the finest level and then evolved with the Schrödinger-Poisson equations, via a fourth-order Runge-Kutta algorithm~\cite{Schwabe2016}. 
The evolution of the N-body particles is tracked inside the Schrödinger-Poisson domain but the source of gravity is the density $\rho = \abs{\psi}^2$ of the wave function. 

Once the halo has formed up to three additional finite-difference levels are added (the level setup is illustrated in  \cref{fig:dens_zoom}) to ensure that the de Broglie wavelength is resolved throughout our simulations. This  yields a total of eight refinement levels with a refinement factor of two at each step, six of which dynamically trace the position of the selected halo.

\section{Simulation Results}
\label{sec:results}

Our zoom-in simulations are related to five halos covering a mass range between $8.5\,\mathrm{g}$ and $225\,\mathrm{g}$ at $\mathcal{N}=17.3$. Analogously to FDM halos, a solitonic core forms in the center of each collapsing inflaton halo surrounded by incoherent granular density fluctuations, as shown in \cref{fig:dens_zoom}.

The solitonic core is identified by its radial density profile~\cite{Schive:2014hza}
\begin{align}
    \rho_\ast(r) \simeq \rho_{\ast,0}\left(1 + 0.091\left(\frac{r}{r_\ast}\right)^2\right)^{-8}\,,
    \label{eq:soliton_profile}
\end{align}
where $r_\ast$ is the (physical) core radius at which the density has dropped to half of its central value
\begin{align}
    \rho_{\ast,0} = 3.3\times 10^{13} \left(\frac{1.546\times 10^{22}\,\mathrm{eV}}{m}\right)^2 \left(\frac{10^{-3}\,l_u}{r_\ast}\right)^4 \frac{m_u}{l_u^3}\,.
    \label{eq:soliton_central_density}
 \end{align}
In \cref{fig:densprofiles} we show radial density profiles centered around the point of maximal density  of the five simulated halos at different $\mathcal{N}$. They are well fitted by \cref{eq:soliton_profile} in the innermost region, transitioning to outer profiles indistinguishable from the results of the corresponding pure N-body simulations. This confirms the hypothesized existence of inflaton stars in the post-inflationary universe~\cite{Niemeyer2019} and demonstrates the validity of the Schrödinger-Vlasov correspondence in the outer parts of the halos. 

\begin{figure}
    \centering
    \includegraphics{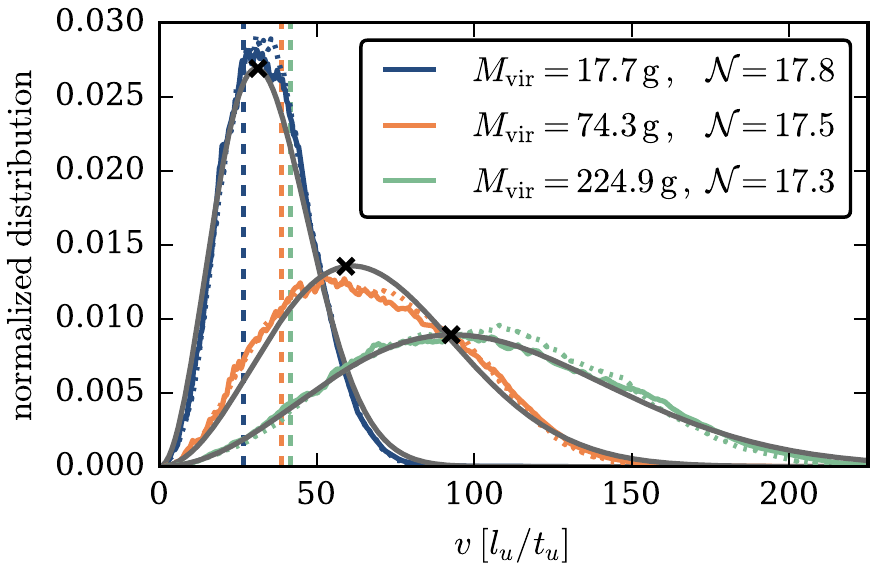}
    \caption{Velocity spectra of three halos inside their respective virial radii for the wave function (solid) and for N-body particles (dotted) in the same region. Fits of Maxwellian distribution functions (cf. \cref{eq:maxwellian}) are displayed in grey. The vertical dashed lines show the soliton velocities $v_\ast$. The crosses mark $v_\mathrm{vir}$ of the respective halos. They closely align with the peaks of the Maxwellian distribution.}
    \label{fig:veldist}
\end{figure}

The velocity distributions of the lowest-, highest- and a medium-mass halo from our sample
are shown in \cref{fig:veldist} for both the wave function $\psi$ within a cube defined by the halo's virial radius $r_\mathrm{vir}$ and for the corresponding pure N-body simulation. The former is given by~\cite{Veltmaat:2018dfz}
\begin{align}
    f(\mathbf{v}) = \frac{1}{N}\abs{\int \mathrm{d}^3\mathbf{x}\exp\left(-im\mathbf{v}\cdot\mathbf{x}/\hbar\right)\psi(\mathbf{x})}^2\,,
\end{align}
where $N$ is a normalization factor. Here, $r_\mathrm{vir}$ defines a sphere around the halo's center with mean density $\Delta_\mathrm{vir}\bar\rho$ and $\Delta_\mathrm{vir}=18\pi^2$ for a matter-dominated universe. The halo's virial mass is then given by $M_\mathrm{vir} = 4\pi/3\Delta_\mathrm{vir}\bar\rho r_\mathrm{vir}^3$ defining its virial velocity as $v_\mathrm{vir} = (GM_\mathrm{vir}/r_\mathrm{vir})^{1/2}$. 

These velocity distributions are almost indistinguishable, again confirming the Schrödinger-Vlasov correspondence, and are well fitted by a Maxwellian distribution
\begin{align}
    f_M(v)\mathrm{d}v = 3\left(\frac{6}{\pi}\right)^{1/2}\frac{v^2}{v_0^3}\exp\left(-\frac{3}{2}\frac{v^2}{v_0^2}\right)\mathrm{d}v\,,
    \label{eq:maxwellian}
\end{align}
with free parameter $v_0$. They peak at $v_\mathrm{vir} = (2/3)^{1/2}v_0$. 

It was previously found that the core radii $r_\ast$ are correlated with the peaks of the velocity distributions of their host halos,
\begin{align} 
    r_\ast = \frac{2\pi}{7.5}\frac{\hbar}{mv_\mathrm{vir}}\,,
    \label{eq:soliton_vel}
\end{align}
once the cores are in virial equilibrium ($2E_\mathrm{kin}+E_\mathrm{pot}=0$) with their surroundings~\cite{Mocz2017,Schwabe2020}. In practice, this means that the soliton velocity satisfies $v_\ast = v_\mathrm{vir}$~\cite{Niemeyer2019_review}. We use \cref{eq:soliton_vel} to compute $v_\ast$ from $r_\ast$; these are displayed as vertical dashed lines in \cref{fig:veldist}. 
While the soliton velocities inside the lower-mass halos agree well with $v_\mathrm{vir}$ at their respective $\mathcal{N}$, $v_\ast$ of the heaviest halo at an earlier time is only half the peak velocity, implying that this core needs to double its mass before reaching virial equilibrium (cf.~\cref{eq:soliton_mass_vvir}).  

We define the mass of a solitonic core as the mass enclosed by $r_\ast$~\cite{Hui2017},
\begin{align}
    M_\ast = 3.9251\frac{\hbar^2}{Gr_\ast m^2}\,.
    \label{eq:soliton_mass_def}
\end{align}
For the density profiles shown in \cref{fig:densprofiles} we find core masses ranging from $0.95\,\mathrm{g}$ to $1.90\,\mathrm{g}$. 
Inserting $r_\ast$ from \cref{eq:soliton_mass_def} into \cref{eq:soliton_vel}, one obtains a relation between the core mass and the virial velocity of the halo:
\begin{align}
    M_\ast = 4.69 \frac{\hbar}{m}\frac{v_\mathrm{vir}}{G}\,.
    \label{eq:soliton_mass_vvir}
\end{align}
Since \cref{eq:soliton_mass_vvir} is only accurate when the soliton radii approach \cref{eq:soliton_vel}
the core masses from the two lighter halos in \cref{fig:veldist} are $15\%$ and $35\%$ smaller than expected from \cref{eq:soliton_mass_vvir}, while the most massive core has only half of its expected mass. 

\begin{figure}
    \centering
    \includegraphics{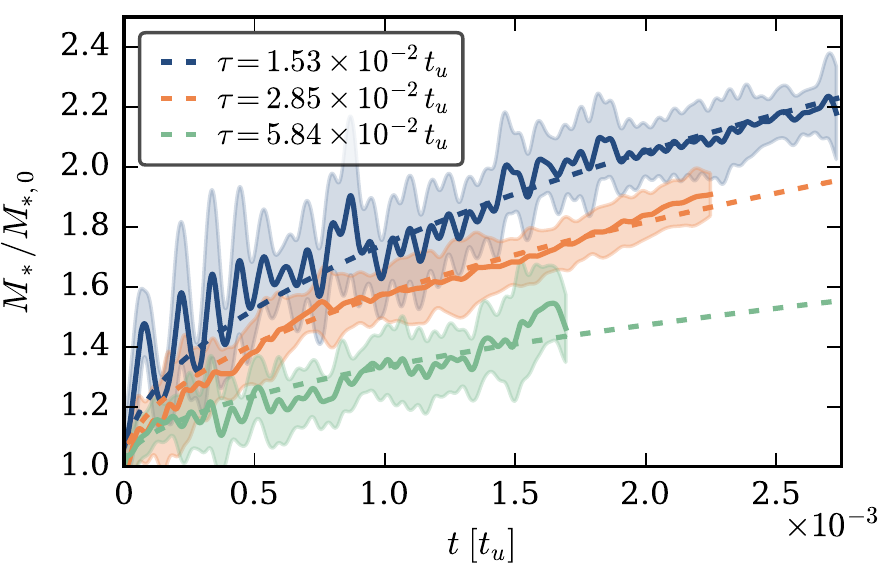}
    \caption{Mass increase of three solitonic cores relative to their mass $M_{\ast,0}$ at formation. Due to the strong oscillations of the solitons the simulation data was smoothed with a Gaussian kernel with a standard deviation of $\sigma = 2\times 10^{-5}\,t_u$. The mean and its $1\sigma$ deviation band are displayed by the solid lines and the shaded regions, respectively. As shown by the dashed lines, the mass growth obeys \cref{eq:mass_growth} with time scale $\tau$ given by \cref{eq:condensation_time}.}
    \label{fig:mass_growth}
\end{figure}

\begin{figure}
    \centering
    \includegraphics{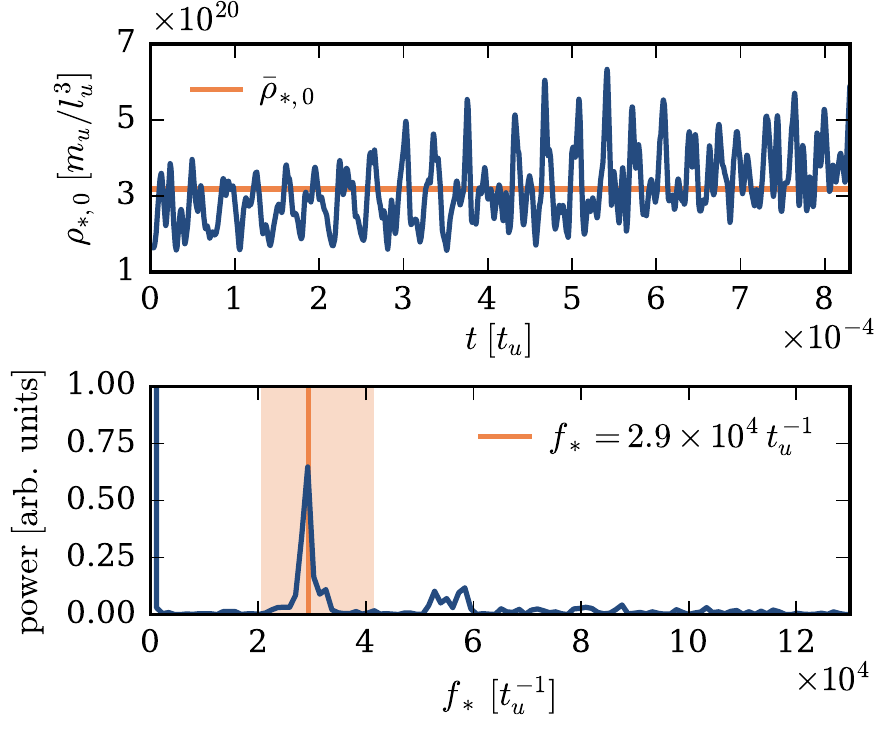}
    \caption{Top: Central soliton density as a function of time. Bottom: Its corresponding power spectrum. Boundaries of the shaded region are determined by the quasi-normal frequencies (see \cref{eq:frequency}) using the maximum and minimum $\rho_{\ast,0}$. The spectrum peaks at the frequency $f_\ast = 2.9\times 10^4\,t_u$ corresponding to $\bar\rho_{\ast,0}$.}
    \label{fig:oscillation}
\end{figure}
\begin{figure*}
    \centering
    \includegraphics[width=\textwidth]{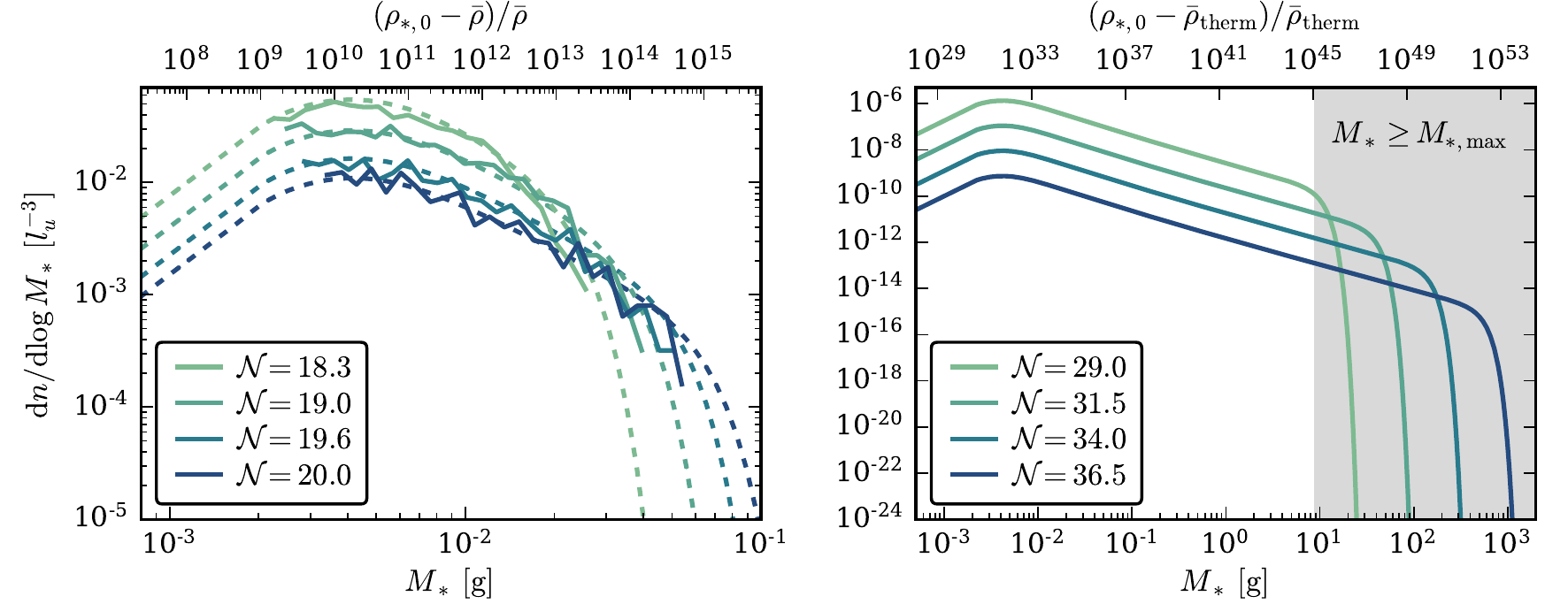}
    \caption{Inflaton star mass function at different number of $e$-folds $\mathcal{N}$ after the end of inflation for an inflaton mass of $m = 6.35\times 10^{-6}\,M_\mathrm{Pl}$. Left: solid lines represent the ISMF derived from the numerically obtained IHMF in Ref.~\cite{Eggemeier2020} while dashed lines display the corresponding mass functions from the Press-Schechter formalism. The upper axis shows the inflaton star overdensity which can be calculated from their mass. Right: ISMF from Press-Schechter theory until thermalization at $T=1\,\mathrm{TeV}$ ($\mathcal{N}=36.5$) with overdensities relative to the mean density $\bar\rho_\mathrm{therm}$ at thermalization. The grey shaded area highlights masses larger than $M_{\ast,\mathrm{max}}$ (see \cref{eq:mass_schwarzschild}). }
    \label{fig:inflatonstarMF}
\end{figure*}

\cref{fig:mass_growth} shows the mass increase of the respective solitonic core inside a low-mass (blue), a medium-mass (orange) and a high-mass halo (green). The soliton masses are
normalized by their initial masses $M_{\ast,0}$ when the radial density profile starts to be well described by the soliton profile defined in \cref{eq:soliton_profile}.
The growth rate depends on the condensation time scale~\cite{Levkov2018}
\begin{align}
    \tau = \frac{0.7\sqrt{2}}{12\pi^3}\left(\frac{m}{\hbar}\right)^3\frac{v_\mathrm{vir}^6}{G^2\rho^2\Lambda}\,,
    \label{eq:condensation_time}
\end{align}
where $\Lambda = \log(mv_\mathrm{vir}r_\mathrm{vir}/\hbar)$ is the Coulomb logarithm and $\rho = \Delta_\mathrm{vir}\bar\rho$ is the mean density of the host halo. 
Computing $\tau$ at soliton formation time, we recover the expected growth rate~\cite{Levkov2018,Eggemeier2019PRD,Jiajun2020}
\begin{align}
    M_\ast  = M_{\ast,0}\left[c\left(\frac{t}{\tau}\right)^{1/2} + 1\right]\,,
    \label{eq:mass_growth}
\end{align}
where $c\simeq 3$ was determined from fitting \cref{eq:mass_growth} to the growth curves.
As is noticeable from \cref{fig:mass_growth} and expected from \cref{eq:condensation_time,eq:mass_growth}, larger halo masses lead to larger values of $\tau$ and thus to a relatively slower mass growth of the solitonic core. It was conjectured in Ref.~\cite{Eggemeier2019PRD} and confirmed in Ref.~\cite{Jiajun2020} that the mass growth proportional to $\sim t^{1/2}$ comes to an end once the soliton is in virial equilibrium with its halo and that it afterwards proceeds with the reduced growth rate $\sim t^{1/8}$. However, our simulations are limited by spatial resolution and it is too computationally expensive to verify this in our setup.

From \cref{fig:mass_growth} we see that strong soliton oscillations cause $M_\ast$ to vary over time by up to $20\%$. 
In agreement with previous studies (cf. Refs.~\cite{Veltmaat:2018dfz,Eggemeier2019PRD,Schwabe2020}) the frequency spectra of the oscillations peak at the quasi-normal mode of the excited solitons~\cite{Guzman2004}
\begin{align}
    f_\ast = 5.2\times 10^4\left(\frac{\bar\rho_{\ast,0}}{10^{21}\,m_u/l_u^3}\right)^{1/2}\,t_u^{-1}\,,
    \label{eq:frequency}
\end{align}
where $\bar\rho_{\ast,0}$ is the time-averaged value of $\rho_{\ast,0}$. A representative frequency analysis is shown in \cref{fig:oscillation}.

\section{Inflaton star mass function} 
\label{sec:inflatonstar_MF}

The mass distribution of inflaton stars can be predicted using the mass distribution of inflaton halos and their radii at different times found in the N-body simulations in Ref.~\cite{Eggemeier2020}.
We compute the expected mass of an inflaton star in each halo with \cref{eq:soliton_mass_vvir}. The resulting inflaton star mass function (ISMF) $\mathrm{d}n/\mathrm{d}\log M_\ast$, defined as the comoving number density of inflaton stars per logarithmic mass interval\footnote{Note that a similar procedure is also applicable to determine the mass distributions of solitonic cores in FDM halos and axion stars in axion miniclusters, respectively.}, is shown in the left panel of \cref{fig:inflatonstarMF} from $\mathcal{N}=18.3$ to $\mathcal{N}=20.0$. Inserting the obtained masses $M_\ast$ into \cref{eq:soliton_mass_def,eq:soliton_central_density} we relate them to their central densities and determine their overdensities relative to the mean density $\bar\rho$. As is evident from the upper axis in the left panel of \cref{fig:inflatonstarMF}, the distribution of  overdensities associated with inflaton stars when $\mathcal{N}=20.0$ peaks at around $10^{10}$ and can reach values as high as $10^{15}$ before dropping super-exponentially.  

The IHMF matches predictions~\cite{Eggemeier2020} from the Press-Schechter formalism~\cite{Press1974} with a sharp-$k$ filter~\cite{Niemeyer2019,Schneider2013,Schneider2015}.  
Assuming that each halo contains an inflaton star with a mass given by \cref{eq:soliton_mass_vvir}, the ISMF retains in general the shape of the Press-Schechter-IHMF  but is shifted to correspondingly smaller masses. As evident from the left panel in \cref{fig:inflatonstarMF}, the corresponding ISMFs 
agree with each other at different $\mathcal{N}$. This allows to extrapolate the ISMF and their distribution of overdensities to even later times in the post-inflationary universe, as shown in the right panel of \cref{fig:inflatonstarMF}. 
Assuming that thermalization occurs at a temperature of $1\,\mathrm{TeV}$, the early matter-dominated epoch lasts for $\mathcal{N}=36.5$ $e$-folds (see Eq. (7) in Ref.~\cite{Eggemeier2020}) allowing typical overdensities at the center of inflaton stars  to be as large as  $10^{32}$ at the end of the matter dominated phase.

An inflaton star, like any boson star, has a maximum possible mass, beyond which its radius approaches the Schwarzschild radius $r_\mathrm{S} = 2GM/c^2$. Applying the uncertainty principle to an inflaton star with maximum momentum $p=mc$ and size $r_\mathrm{S}$, one obtains an estimate for the upper mass limit~\cite{Liebling2012,Seidel1990,Padilla:2021zgm}
\begin{align}
    M_{\ast,\mathrm{max}} = \frac{1}{2}\frac{\hbar}{m}\frac{c}{G}\,,
    \label{eq:mass_schwarzschild}
\end{align}
above which an inflaton star becomes gravitationally unstable, potentially dispersing or collapsing into a black hole depending on the details of its prior evolution~\cite{Seidel1990,Helfer2017}. 
As is shown in the right panel of \cref{fig:inflatonstarMF}, this condition is met for $\mathcal{N}\geq 29$ $e$-folds after the end of inflation.

Integrating the ISMF yields an estimate for the mass $M_{\ast,\mathrm{tot}}$ per unit volume that is attributed to the sum of the masses of all inflaton stars, 
\begin{align}
    M_{\ast,\mathrm{tot}} = \int_0^{M_{\ast,\mathrm{max}}} \frac{\mathrm{d}n}{\mathrm{d}\log M_\ast}\,\mathrm{d} M_\ast\,.
    \label{eq:integral_ISMF}
\end{align}
While the fraction of mass that is bound in inflaton halos rapidly converges to $70\%$ already at $\mathcal{N}=25$, we find that the mass fraction of inflaton stars decreases linearly with scale factor from $10^{-5}$ at $\mathcal{N}=20.0$ to $10^{-12}$ at $\mathcal{N}=36.5$. This is reasonable since overall $\mathrm{d}n/\mathrm{d}\log M_\ast\sim a^{-1}$ while the cutoff at the high-mass end of the ISMF (see right panel of \cref{fig:inflatonstarMF}), which effectively acts as the upper limit in \cref{eq:integral_ISMF} when $M_{\ast,\mathrm{max}}$ is not yet reached, increases only weakly with $a$. 

Setting the lower limit in the integral to $M_{\ast,\mathrm{max}}$ and integrating to $M_\star \rightarrow \infty$ yields the integrated mass of inflaton stars with $M\ge M_{\ast,\mathrm{max}}$. The maximal fraction of $3\times 10^{-11}$ is reached at $\mathcal{N}=30$, afterwards it decreases. 
Importantly, this quantity is not the mass fraction of black holes since (i) not every unstable inflaton star collapses into a black hole and (ii) this approach does not capture black hole evolution via accretion, mergers or evaporation. With maximal possible masses of only $1\,\mathrm{kg}$ assuming thermalization at $T=1\,\mathrm{TeV}$, any such black holes would evaporate in less than $10^{-18}\,\mathrm{s}$~\cite{Hooper2019}.

\section{Conclusions}
\label{sec:conclusions}

Using a hybrid method  combining an N-body scheme at large scales with a finite-difference solver for the Schrödinger-Poisson equations at small scales we have performed highly refined simulations of the effective matter-dominated era in the post-inflationary universe. 
This allows us to extend the N-body simulations in Ref.~\cite{Eggemeier2020} to much smaller length scales by zooming into selected halos to properly resolve their interior structure. Taking a sample of five halos with masses ranging from $8.5\,\mathrm{g}$ to $225\,\mathrm{g}$ at $\mathcal{N}=17.3$ $e$-folds after the end of inflation, we confirm the existence of solitonic cores (inflaton stars) in the very early universe.

In agreement with their late-time  analogs in FDM halos~\cite{Schive:2014dra,Schive:2014hza,Veltmaat:2018dfz} and axion miniclusters~\cite{Eggemeier2019PRD}, we find that inflaton stars are subject to strong quasi-normal oscillations and that their masses after formation  increase as $t^{1/2}$. The radial density profiles in the outer regions of the simulated halos and the velocity distribution inside them are consistent with their counterparts in pure N-body simulations, confirming the validity of the Schrödinger-Vlasov correspondence on scales larger than the de Broglie wavelength.

Making use of the IHMF from our previous N-body simulations~\cite{Eggemeier2020} and the Press-Schechter formalism we predict the mass distribution of inflaton stars in the post-inflationary universe until thermalization. At $\mathcal{N}=20$ the mass contained in inflaton stars constitutes $\sim 10^{-5}$ of the total mass while $70\%$ of the mass is bound in inflaton halos. Since halo masses increase over time, the inflaton star masses rise according to \cref{eq:soliton_mass_vvir}. Eventually, they reach the upper mass limit $M_{\ast,\mathrm{max}}$, suggesting that they can collapse into a PBH~\cite{Carr1974,Carr1975,Green1997,Green2001}. 
However, the mass fraction of collapsed inflaton stars is less than $10^{-10}$.

Given their masses, all these black holes evaporate prior to Big Bang Nucleosynthesis. Any particle coupled to gravity whose mass is lower than the current Hawking temperature will be radiated by a decaying black hole~\cite{Hawking1975}.  This includes any stable particles within the overall spectrum of ultra-high energy physics, whether or not they couple directly to the inflaton. As a consequence, a decaying PBH population  creates an alternative mechanism to generate dark matter~\cite{Fujita2014,Lennon2018,Allahverdi2018,Morrison2019,Masina2020,Cheek2021}. Conversely, massless states could contribute to dark radiation~\cite{Hooper2019,Masina2020}. This primordial black hole formation mechanism was discussed in a recent paper by Padilla {\em et al.}~\cite{Padilla:2021zgm} and a full analysis  (and its consistency with other formation mechanisms posited for the post-inflationary universe~\cite{Khlopov1985,Harada2016,Hidalgo2017,Carr2018,Martin2020}) is an obvious avenue for future work. Likewise, detailed studies of the onset and ultimate consequences of solitons exceeding the bound of \cref{eq:mass_schwarzschild} is an important topic, which will require fully relativistic simulations~\cite{Helfer2017}.

Whether or not black holes form after the inflaton stars become unstable,  the formation and interaction of inflaton halos and inflaton stars will source gravitational waves~\cite{Jedamzik2010_grav}.  The resulting gravitational radiation would contribute to a (high frequency) stochastic gravitational wave background that may be detectable with future experiments~\cite{Aggarwal2020} and this is likewise  a fruitful topic for further analysis.

Observational constraints suggest that the inflationary potential is sub-quadratic for large field values, and  non-quadratic contributions can lead to a phase of parametric resonance immediately after the end of inflation. Resonance reprocesses the initial spectrum and this could significantly accelerate the formation of gravitationally bound structures.  Thus, one can expect a rich phenomenology of inflaton halos and inflaton stars prior to thermalization in a broad class of inflationary scenarios, provided the reheating temperature is sufficiently low. 
Moreover, if the inflaton has (weak) self-couplings during this phase, the Schr\"{o}dinger-Poisson equations will need to be generalized to the Gross–Pitaevskii equations to capture them.  Separately, there are likely analogues of the resonant decay of axion stars into photons~\cite{Hertzberg2018,Levkov2020} for inflaton stars coupled to external fields. 

In this paper we have taken steps towards building a full understanding of the rich nonlinear dynamics that can occur in the post-inflationary universe. These arise in the simplest inflationary models, but  there is still much left to understand. This includes the possibility of black hole formation and the consequences of both the inflaton self-coupling, its interactions with other fields, and the processes responsible for thermalizing the universe. 

\vspace*{5mm}
\section*{Acknowledgements}
We thank  Mateja Gosenca,  Peter Hayman, Shaun Hotchkiss, and Emily Kendall for useful discussions. Computations described in this work were performed with resources provided by the North-German Supercomputing Alliance (HLRN). We acknowledge the yt toolkit~\cite{Turk2011} that was used for the analysis of numerical data. RE acknowledges support from the Marsden Fund of the Royal Society of New Zealand. JCN acknowledges a Julius von Haast Fellowship Award provided by the New Zealand Ministry of Business, Innovation and Employment and administered by the Royal Society of New Zealand.  BS acknowledges support by the Deutsche Forschungsgemeinschaft and by grant PGC2018-095328-B-I00(FEDER/Agencia estatal de investigaci\'on).

\bibliography{refs}

\begin{thebibliography}{73}%
\makeatletter
\providecommand \@ifxundefined [1]{%
 \@ifx{#1\undefined}
}%
\providecommand \@ifnum [1]{%
 \ifnum #1\expandafter \@firstoftwo
 \else \expandafter \@secondoftwo
 \fi
}%
\providecommand \@ifx [1]{%
 \ifx #1\expandafter \@firstoftwo
 \else \expandafter \@secondoftwo
 \fi
}%
\providecommand \natexlab [1]{#1}%
\providecommand \enquote  [1]{``#1''}%
\providecommand \bibnamefont  [1]{#1}%
\providecommand \bibfnamefont [1]{#1}%
\providecommand \citenamefont [1]{#1}%
\providecommand \href@noop [0]{\@secondoftwo}%
\providecommand \href [0]{\begingroup \@sanitize@url \@href}%
\providecommand \@href[1]{\@@startlink{#1}\@@href}%
\providecommand \@@href[1]{\endgroup#1\@@endlink}%
\providecommand \@sanitize@url [0]{\catcode `\\12\catcode `\$12\catcode
  `\&12\catcode `\#12\catcode `\^12\catcode `\_12\catcode `\%12\relax}%
\providecommand \@@startlink[1]{}%
\providecommand \@@endlink[0]{}%
\providecommand \url  [0]{\begingroup\@sanitize@url \@url }%
\providecommand \@url [1]{\endgroup\@href {#1}{\urlprefix }}%
\providecommand \urlprefix  [0]{URL }%
\providecommand \Eprint [0]{\href }%
\providecommand \doibase [0]{http://dx.doi.org/}%
\providecommand \selectlanguage [0]{\@gobble}%
\providecommand \bibinfo  [0]{\@secondoftwo}%
\providecommand \bibfield  [0]{\@secondoftwo}%
\providecommand \translation [1]{[#1]}%
\providecommand \BibitemOpen [0]{}%
\providecommand \bibitemStop [0]{}%
\providecommand \bibitemNoStop [0]{.\EOS\space}%
\providecommand \EOS [0]{\spacefactor3000\relax}%
\providecommand \BibitemShut  [1]{\csname bibitem#1\endcsname}%
\let\auto@bib@innerbib\@empty
\bibitem [{\citenamefont {Starobinsky}(1980)}]{Starobinsky1980}%
  \BibitemOpen
  \bibfield  {author} {\bibinfo {author} {\bibfnamefont {A.}~\bibnamefont
  {Starobinsky}},\ }\href {\doibase
  https://doi.org/10.1016/0370-2693(80)90670-X} {\bibfield  {journal} {\bibinfo
   {journal} {Physics Letters B}\ }\textbf {\bibinfo {volume} {91}},\ \bibinfo
  {pages} {99 } (\bibinfo {year} {1980})}\BibitemShut {NoStop}%
\bibitem [{\citenamefont {Guth}(1981)}]{Guth1981}%
  \BibitemOpen
  \bibfield  {author} {\bibinfo {author} {\bibfnamefont {A.~H.}\ \bibnamefont
  {Guth}},\ }\href {\doibase 10.1103/PhysRevD.23.347} {\bibfield  {journal}
  {\bibinfo  {journal} {Phys. Rev. D}\ }\textbf {\bibinfo {volume} {23}},\
  \bibinfo {pages} {347} (\bibinfo {year} {1981})}\BibitemShut {NoStop}%
\bibitem [{\citenamefont {Linde}(1982)}]{Linde1982}%
  \BibitemOpen
  \bibfield  {author} {\bibinfo {author} {\bibfnamefont {A.}~\bibnamefont
  {Linde}},\ }\href {\doibase https://doi.org/10.1016/0370-2693(82)91219-9}
  {\bibfield  {journal} {\bibinfo  {journal} {Physics Letters B}\ }\textbf
  {\bibinfo {volume} {108}},\ \bibinfo {pages} {389 } (\bibinfo {year}
  {1982})}\BibitemShut {NoStop}%
\bibitem [{\citenamefont {Linde}(1983)}]{Linde1983}%
  \BibitemOpen
  \bibfield  {author} {\bibinfo {author} {\bibfnamefont {A.}~\bibnamefont
  {Linde}},\ }\href {\doibase https://doi.org/10.1016/0370-2693(83)90837-7}
  {\bibfield  {journal} {\bibinfo  {journal} {Physics Letters B}\ }\textbf
  {\bibinfo {volume} {129}},\ \bibinfo {pages} {177 } (\bibinfo {year}
  {1983})}\BibitemShut {NoStop}%
\bibitem [{\citenamefont {Traschen}\ and\ \citenamefont
  {Brandenberger}(1990)}]{Traschen:1990sw}%
  \BibitemOpen
  \bibfield  {author} {\bibinfo {author} {\bibfnamefont {J.~H.}\ \bibnamefont
  {Traschen}}\ and\ \bibinfo {author} {\bibfnamefont {R.~H.}\ \bibnamefont
  {Brandenberger}},\ }\href {\doibase 10.1103/PhysRevD.42.2491} {\bibfield
  {journal} {\bibinfo  {journal} {Phys. Rev. D}\ }\textbf {\bibinfo {volume}
  {42}},\ \bibinfo {pages} {2491} (\bibinfo {year} {1990})}\BibitemShut
  {NoStop}%
\bibitem [{\citenamefont {Shtanov}\ \emph {et~al.}(1995)\citenamefont
  {Shtanov}, \citenamefont {Traschen},\ and\ \citenamefont
  {Brandenberger}}]{Shtanov:1994ce}%
  \BibitemOpen
  \bibfield  {author} {\bibinfo {author} {\bibfnamefont {Y.}~\bibnamefont
  {Shtanov}}, \bibinfo {author} {\bibfnamefont {J.~H.}\ \bibnamefont
  {Traschen}}, \ and\ \bibinfo {author} {\bibfnamefont {R.~H.}\ \bibnamefont
  {Brandenberger}},\ }\href {\doibase 10.1103/PhysRevD.51.5438} {\bibfield
  {journal} {\bibinfo  {journal} {Phys. Rev.}\ }\textbf {\bibinfo {volume}
  {D51}},\ \bibinfo {pages} {5438} (\bibinfo {year} {1995})},\ \Eprint
  {http://arxiv.org/abs/hep-ph/9407247} {arXiv:hep-ph/9407247 [hep-ph]}
  \BibitemShut {NoStop}%
\bibitem [{\citenamefont {Kofman}\ \emph {et~al.}(1997)\citenamefont {Kofman},
  \citenamefont {Linde},\ and\ \citenamefont {Starobinsky}}]{Kofman1997}%
  \BibitemOpen
  \bibfield  {author} {\bibinfo {author} {\bibfnamefont {L.}~\bibnamefont
  {Kofman}}, \bibinfo {author} {\bibfnamefont {A.}~\bibnamefont {Linde}}, \
  and\ \bibinfo {author} {\bibfnamefont {A.~A.}\ \bibnamefont {Starobinsky}},\
  }\href {\doibase 10.1103/PhysRevD.56.3258} {\bibfield  {journal} {\bibinfo
  {journal} {Phys. Rev. D}\ }\textbf {\bibinfo {volume} {56}},\ \bibinfo
  {pages} {3258} (\bibinfo {year} {1997})}\BibitemShut {NoStop}%
\bibitem [{\citenamefont {Lozanov}\ and\ \citenamefont
  {Amin}(2017)}]{Lozanov:2016hid}%
  \BibitemOpen
  \bibfield  {author} {\bibinfo {author} {\bibfnamefont {K.~D.}\ \bibnamefont
  {Lozanov}}\ and\ \bibinfo {author} {\bibfnamefont {M.~A.}\ \bibnamefont
  {Amin}},\ }\href {\doibase 10.1103/PhysRevLett.119.061301} {\bibfield
  {journal} {\bibinfo  {journal} {Phys. Rev. Lett.}\ }\textbf {\bibinfo
  {volume} {119}},\ \bibinfo {pages} {061301} (\bibinfo {year} {2017})},\
  \Eprint {http://arxiv.org/abs/1608.01213} {arXiv:1608.01213 [astro-ph.CO]}
  \BibitemShut {NoStop}%
\bibitem [{\citenamefont {Gleiser}(1994)}]{Gleiser:1993pt}%
  \BibitemOpen
  \bibfield  {author} {\bibinfo {author} {\bibfnamefont {M.}~\bibnamefont
  {Gleiser}},\ }\href {\doibase 10.1103/PhysRevD.49.2978} {\bibfield  {journal}
  {\bibinfo  {journal} {Phys. Rev. D}\ }\textbf {\bibinfo {volume} {49}},\
  \bibinfo {pages} {2978} (\bibinfo {year} {1994})},\ \Eprint
  {http://arxiv.org/abs/hep-ph/9308279} {arXiv:hep-ph/9308279} \BibitemShut
  {NoStop}%
\bibitem [{\citenamefont {Copeland}\ \emph {et~al.}(1995)\citenamefont
  {Copeland}, \citenamefont {Gleiser},\ and\ \citenamefont
  {Muller}}]{Copeland:1995fq}%
  \BibitemOpen
  \bibfield  {author} {\bibinfo {author} {\bibfnamefont {E.~J.}\ \bibnamefont
  {Copeland}}, \bibinfo {author} {\bibfnamefont {M.}~\bibnamefont {Gleiser}}, \
  and\ \bibinfo {author} {\bibfnamefont {H.~R.}\ \bibnamefont {Muller}},\
  }\href {\doibase 10.1103/PhysRevD.52.1920} {\bibfield  {journal} {\bibinfo
  {journal} {Phys. Rev. D}\ }\textbf {\bibinfo {volume} {52}},\ \bibinfo
  {pages} {1920} (\bibinfo {year} {1995})},\ \Eprint
  {http://arxiv.org/abs/hep-ph/9503217} {arXiv:hep-ph/9503217} \BibitemShut
  {NoStop}%
\bibitem [{\citenamefont {Amin}\ \emph {et~al.}(2010)\citenamefont {Amin},
  \citenamefont {Easther},\ and\ \citenamefont {Finkel}}]{Amin2010}%
  \BibitemOpen
  \bibfield  {author} {\bibinfo {author} {\bibfnamefont {M.~A.}\ \bibnamefont
  {Amin}}, \bibinfo {author} {\bibfnamefont {R.}~\bibnamefont {Easther}}, \
  and\ \bibinfo {author} {\bibfnamefont {H.}~\bibnamefont {Finkel}},\ }\href
  {\doibase 10.1088/1475-7516/2010/12/001} {\bibfield  {journal} {\bibinfo
  {journal} {Journal of Cosmology and Astroparticle Physics}\ }\textbf
  {\bibinfo {volume} {2010}},\ \bibinfo {pages} {001} (\bibinfo {year}
  {2010})}\BibitemShut {NoStop}%
\bibitem [{\citenamefont {Amin}\ \emph {et~al.}(2012)\citenamefont {Amin},
  \citenamefont {Easther}, \citenamefont {Finkel}, \citenamefont {Flauger},\
  and\ \citenamefont {Hertzberg}}]{Amin2012}%
  \BibitemOpen
  \bibfield  {author} {\bibinfo {author} {\bibfnamefont {M.~A.}\ \bibnamefont
  {Amin}}, \bibinfo {author} {\bibfnamefont {R.}~\bibnamefont {Easther}},
  \bibinfo {author} {\bibfnamefont {H.}~\bibnamefont {Finkel}}, \bibinfo
  {author} {\bibfnamefont {R.}~\bibnamefont {Flauger}}, \ and\ \bibinfo
  {author} {\bibfnamefont {M.~P.}\ \bibnamefont {Hertzberg}},\ }\href {\doibase
  10.1103/PhysRevLett.108.241302} {\bibfield  {journal} {\bibinfo  {journal}
  {Phys. Rev. Lett.}\ }\textbf {\bibinfo {volume} {108}},\ \bibinfo {pages}
  {241302} (\bibinfo {year} {2012})}\BibitemShut {NoStop}%
\bibitem [{\citenamefont {Lozanov}\ and\ \citenamefont
  {Amin}(2018)}]{Lozanov:2017hjm}%
  \BibitemOpen
  \bibfield  {author} {\bibinfo {author} {\bibfnamefont {K.~D.}\ \bibnamefont
  {Lozanov}}\ and\ \bibinfo {author} {\bibfnamefont {M.~A.}\ \bibnamefont
  {Amin}},\ }\href {\doibase 10.1103/PhysRevD.97.023533} {\bibfield  {journal}
  {\bibinfo  {journal} {Phys. Rev. D}\ }\textbf {\bibinfo {volume} {97}},\
  \bibinfo {pages} {023533} (\bibinfo {year} {2018})},\ \Eprint
  {http://arxiv.org/abs/1710.06851} {arXiv:1710.06851 [astro-ph.CO]}
  \BibitemShut {NoStop}%
\bibitem [{\citenamefont {Easther}\ \emph {et~al.}(2011)\citenamefont
  {Easther}, \citenamefont {Flauger},\ and\ \citenamefont
  {Gilmore}}]{Easther2010}%
  \BibitemOpen
  \bibfield  {author} {\bibinfo {author} {\bibfnamefont {R.}~\bibnamefont
  {Easther}}, \bibinfo {author} {\bibfnamefont {R.}~\bibnamefont {Flauger}}, \
  and\ \bibinfo {author} {\bibfnamefont {J.~B.}\ \bibnamefont {Gilmore}},\
  }\href {\doibase 10.1088/1475-7516/2011/04/027} {\bibfield  {journal}
  {\bibinfo  {journal} {JCAP}\ }\textbf {\bibinfo {volume} {1104}},\ \bibinfo
  {pages} {027} (\bibinfo {year} {2011})},\ \Eprint
  {http://arxiv.org/abs/1003.3011} {arXiv:1003.3011 [astro-ph.CO]} \BibitemShut
  {NoStop}%
\bibitem [{\citenamefont {Jedamzik}\ \emph
  {et~al.}(2010{\natexlab{a}})\citenamefont {Jedamzik}, \citenamefont
  {Lemoine},\ and\ \citenamefont {Martin}}]{Jedamzik:2010hq}%
  \BibitemOpen
  \bibfield  {author} {\bibinfo {author} {\bibfnamefont {K.}~\bibnamefont
  {Jedamzik}}, \bibinfo {author} {\bibfnamefont {M.}~\bibnamefont {Lemoine}}, \
  and\ \bibinfo {author} {\bibfnamefont {J.}~\bibnamefont {Martin}},\ }\href
  {\doibase 10.1088/1475-7516/2010/09/034} {\bibfield  {journal} {\bibinfo
  {journal} {Journal of Cosmology and Astroparticle Physics}\ }\textbf
  {\bibinfo {volume} {2010}},\ \bibinfo {pages} {034} (\bibinfo {year}
  {2010}{\natexlab{a}})}\BibitemShut {NoStop}%
\bibitem [{\citenamefont {Musoke}\ \emph {et~al.}(2020)\citenamefont {Musoke},
  \citenamefont {Hotchkiss},\ and\ \citenamefont {Easther}}]{Musoke2019}%
  \BibitemOpen
  \bibfield  {author} {\bibinfo {author} {\bibfnamefont {N.}~\bibnamefont
  {Musoke}}, \bibinfo {author} {\bibfnamefont {S.}~\bibnamefont {Hotchkiss}}, \
  and\ \bibinfo {author} {\bibfnamefont {R.}~\bibnamefont {Easther}},\ }\href
  {\doibase 10.1103/PhysRevLett.124.061301} {\bibfield  {journal} {\bibinfo
  {journal} {Phys. Rev. Lett.}\ }\textbf {\bibinfo {volume} {124}},\ \bibinfo
  {pages} {061301} (\bibinfo {year} {2020})},\ \Eprint
  {http://arxiv.org/abs/1909.11678} {arXiv:1909.11678 [astro-ph.CO]}
  \BibitemShut {NoStop}%
\bibitem [{\citenamefont {Niemeyer}\ and\ \citenamefont
  {Easther}(2020)}]{Niemeyer2019}%
  \BibitemOpen
  \bibfield  {author} {\bibinfo {author} {\bibfnamefont {J.~C.}\ \bibnamefont
  {Niemeyer}}\ and\ \bibinfo {author} {\bibfnamefont {R.}~\bibnamefont
  {Easther}},\ }\href {\doibase 10.1088/1475-7516/2020/07/030} {\bibfield
  {journal} {\bibinfo  {journal} {Journal of Cosmology and Astroparticle
  Physics}\ }\textbf {\bibinfo {volume} {2020}},\ \bibinfo {pages} {030}
  (\bibinfo {year} {2020})}\BibitemShut {NoStop}%
\bibitem [{\citenamefont {Eggemeier}\ \emph {et~al.}(2021)\citenamefont
  {Eggemeier}, \citenamefont {Niemeyer},\ and\ \citenamefont
  {Easther}}]{Eggemeier2020}%
  \BibitemOpen
  \bibfield  {author} {\bibinfo {author} {\bibfnamefont {B.}~\bibnamefont
  {Eggemeier}}, \bibinfo {author} {\bibfnamefont {J.~C.}\ \bibnamefont
  {Niemeyer}}, \ and\ \bibinfo {author} {\bibfnamefont {R.}~\bibnamefont
  {Easther}},\ }\href {\doibase 10.1103/PhysRevD.103.063525} {\bibfield
  {journal} {\bibinfo  {journal} {Phys. Rev. D}\ }\textbf {\bibinfo {volume}
  {103}},\ \bibinfo {pages} {063525} (\bibinfo {year} {2021})}\BibitemShut
  {NoStop}%
\bibitem [{\citenamefont {Liddle}\ and\ \citenamefont
  {Leach}(2003)}]{Liddle2003}%
  \BibitemOpen
  \bibfield  {author} {\bibinfo {author} {\bibfnamefont {A.~R.}\ \bibnamefont
  {Liddle}}\ and\ \bibinfo {author} {\bibfnamefont {S.~M.}\ \bibnamefont
  {Leach}},\ }\href {\doibase 10.1103/PhysRevD.68.103503} {\bibfield  {journal}
  {\bibinfo  {journal} {Phys. Rev. D}\ }\textbf {\bibinfo {volume} {68}},\
  \bibinfo {pages} {103503} (\bibinfo {year} {2003})},\ \Eprint
  {http://arxiv.org/abs/astro-ph/0305263} {arXiv:astro-ph/0305263} \BibitemShut
  {NoStop}%
\bibitem [{\citenamefont {Adshead}\ and\ \citenamefont
  {Easther}(2008)}]{Adshead2008}%
  \BibitemOpen
  \bibfield  {author} {\bibinfo {author} {\bibfnamefont {P.}~\bibnamefont
  {Adshead}}\ and\ \bibinfo {author} {\bibfnamefont {R.}~\bibnamefont
  {Easther}},\ }\href {\doibase 10.1088/1475-7516/2008/10/047} {\bibfield
  {journal} {\bibinfo  {journal} {JCAP}\ }\textbf {\bibinfo {volume} {10}},\
  \bibinfo {pages} {047} (\bibinfo {year} {2008})},\ \Eprint
  {http://arxiv.org/abs/0802.3898} {arXiv:0802.3898 [astro-ph]} \BibitemShut
  {NoStop}%
\bibitem [{\citenamefont {Adshead}\ \emph {et~al.}(2011)\citenamefont
  {Adshead}, \citenamefont {Easther}, \citenamefont {Pritchard},\ and\
  \citenamefont {Loeb}}]{Adshead2011}%
  \BibitemOpen
  \bibfield  {author} {\bibinfo {author} {\bibfnamefont {P.}~\bibnamefont
  {Adshead}}, \bibinfo {author} {\bibfnamefont {R.}~\bibnamefont {Easther}},
  \bibinfo {author} {\bibfnamefont {J.}~\bibnamefont {Pritchard}}, \ and\
  \bibinfo {author} {\bibfnamefont {A.}~\bibnamefont {Loeb}},\ }\href {\doibase
  10.1088/1475-7516/2011/02/021} {\bibfield  {journal} {\bibinfo  {journal}
  {JCAP}\ }\textbf {\bibinfo {volume} {02}},\ \bibinfo {pages} {021} (\bibinfo
  {year} {2011})},\ \Eprint {http://arxiv.org/abs/1007.3748} {arXiv:1007.3748
  [astro-ph.CO]} \BibitemShut {NoStop}%
\bibitem [{\citenamefont {Hu}\ \emph {et~al.}(2000)\citenamefont {Hu},
  \citenamefont {Barkana},\ and\ \citenamefont {Gruzinov}}]{Hu2000}%
  \BibitemOpen
  \bibfield  {author} {\bibinfo {author} {\bibfnamefont {W.}~\bibnamefont
  {Hu}}, \bibinfo {author} {\bibfnamefont {R.}~\bibnamefont {Barkana}}, \ and\
  \bibinfo {author} {\bibfnamefont {A.}~\bibnamefont {Gruzinov}},\ }\href
  {\doibase 10.1103/PhysRevLett.85.1158} {\bibfield  {journal} {\bibinfo
  {journal} {Phys. Rev. Lett.}\ }\textbf {\bibinfo {volume} {85}},\ \bibinfo
  {pages} {1158} (\bibinfo {year} {2000})}\BibitemShut {NoStop}%
\bibitem [{\citenamefont {Schwabe}\ \emph {et~al.}(2016)\citenamefont
  {Schwabe}, \citenamefont {Niemeyer},\ and\ \citenamefont
  {Engels}}]{Schwabe2016}%
  \BibitemOpen
  \bibfield  {author} {\bibinfo {author} {\bibfnamefont {B.}~\bibnamefont
  {Schwabe}}, \bibinfo {author} {\bibfnamefont {J.~C.}\ \bibnamefont
  {Niemeyer}}, \ and\ \bibinfo {author} {\bibfnamefont {J.~F.}\ \bibnamefont
  {Engels}},\ }\href {\doibase 10.1103/PhysRevD.94.043513} {\bibfield
  {journal} {\bibinfo  {journal} {Phys. Rev. D}\ }\textbf {\bibinfo {volume}
  {94}},\ \bibinfo {pages} {043513} (\bibinfo {year} {2016})}\BibitemShut
  {NoStop}%
\bibitem [{\citenamefont {Niemeyer}(2020)}]{Niemeyer2019_review}%
  \BibitemOpen
  \bibfield  {author} {\bibinfo {author} {\bibfnamefont {J.~C.}\ \bibnamefont
  {Niemeyer}},\ }\href {\doibase https://doi.org/10.1016/j.ppnp.2020.103787}
  {\bibfield  {journal} {\bibinfo  {journal} {Progress in Particle and Nuclear
  Physics}\ }\textbf {\bibinfo {volume} {113}},\ \bibinfo {pages} {103787}
  (\bibinfo {year} {2020})}\BibitemShut {NoStop}%
\bibitem [{\citenamefont {{Widrow}}\ and\ \citenamefont
  {{Kaiser}}(1993)}]{Widrow1993}%
  \BibitemOpen
  \bibfield  {author} {\bibinfo {author} {\bibfnamefont {L.~M.}\ \bibnamefont
  {{Widrow}}}\ and\ \bibinfo {author} {\bibfnamefont {N.}~\bibnamefont
  {{Kaiser}}},\ }\href {\doibase 10.1086/187073} {\bibfield  {journal}
  {\bibinfo  {journal} {\apjl}\ }\textbf {\bibinfo {volume} {416}},\ \bibinfo
  {pages} {L71} (\bibinfo {year} {1993})}\BibitemShut {NoStop}%
\bibitem [{\citenamefont {Uhlemann}\ \emph {et~al.}(2014)\citenamefont
  {Uhlemann}, \citenamefont {Kopp},\ and\ \citenamefont
  {Haugg}}]{Uhlemann2014}%
  \BibitemOpen
  \bibfield  {author} {\bibinfo {author} {\bibfnamefont {C.}~\bibnamefont
  {Uhlemann}}, \bibinfo {author} {\bibfnamefont {M.}~\bibnamefont {Kopp}}, \
  and\ \bibinfo {author} {\bibfnamefont {T.}~\bibnamefont {Haugg}},\ }\href
  {\doibase 10.1103/PhysRevD.90.023517} {\bibfield  {journal} {\bibinfo
  {journal} {Phys. Rev. D}\ }\textbf {\bibinfo {volume} {90}},\ \bibinfo
  {pages} {023517} (\bibinfo {year} {2014})}\BibitemShut {NoStop}%
\bibitem [{\citenamefont {Schive}\ \emph
  {et~al.}(2014{\natexlab{a}})\citenamefont {Schive}, \citenamefont {Chiueh},\
  and\ \citenamefont {Broadhurst}}]{Schive:2014dra}%
  \BibitemOpen
  \bibfield  {author} {\bibinfo {author} {\bibfnamefont {H.-Y.}\ \bibnamefont
  {Schive}}, \bibinfo {author} {\bibfnamefont {T.}~\bibnamefont {Chiueh}}, \
  and\ \bibinfo {author} {\bibfnamefont {T.}~\bibnamefont {Broadhurst}},\
  }\href {\doibase 10.1038/nphys2996} {\bibfield  {journal} {\bibinfo
  {journal} {Nature Phys.}\ }\textbf {\bibinfo {volume} {10}},\ \bibinfo
  {pages} {496} (\bibinfo {year} {2014}{\natexlab{a}})},\ \Eprint
  {http://arxiv.org/abs/1406.6586} {arXiv:1406.6586 [astro-ph.GA]} \BibitemShut
  {NoStop}%
\bibitem [{\citenamefont {Schive}\ \emph
  {et~al.}(2014{\natexlab{b}})\citenamefont {Schive}, \citenamefont {Liao},
  \citenamefont {Woo}, \citenamefont {Wong}, \citenamefont {Chiueh},
  \citenamefont {Broadhurst},\ and\ \citenamefont {Hwang}}]{Schive:2014hza}%
  \BibitemOpen
  \bibfield  {author} {\bibinfo {author} {\bibfnamefont {H.-Y.}\ \bibnamefont
  {Schive}}, \bibinfo {author} {\bibfnamefont {M.-H.}\ \bibnamefont {Liao}},
  \bibinfo {author} {\bibfnamefont {T.-P.}\ \bibnamefont {Woo}}, \bibinfo
  {author} {\bibfnamefont {S.-K.}\ \bibnamefont {Wong}}, \bibinfo {author}
  {\bibfnamefont {T.}~\bibnamefont {Chiueh}}, \bibinfo {author} {\bibfnamefont
  {T.}~\bibnamefont {Broadhurst}}, \ and\ \bibinfo {author} {\bibfnamefont
  {W.-Y.~P.}\ \bibnamefont {Hwang}},\ }\href {\doibase
  10.1103/PhysRevLett.113.261302} {\bibfield  {journal} {\bibinfo  {journal}
  {Phys. Rev. Lett.}\ }\textbf {\bibinfo {volume} {113}},\ \bibinfo {pages}
  {261302} (\bibinfo {year} {2014}{\natexlab{b}})}\BibitemShut {NoStop}%
\bibitem [{\citenamefont {Veltmaat}\ \emph {et~al.}(2018)\citenamefont
  {Veltmaat}, \citenamefont {Niemeyer},\ and\ \citenamefont
  {Schwabe}}]{Veltmaat:2018dfz}%
  \BibitemOpen
  \bibfield  {author} {\bibinfo {author} {\bibfnamefont {J.}~\bibnamefont
  {Veltmaat}}, \bibinfo {author} {\bibfnamefont {J.~C.}\ \bibnamefont
  {Niemeyer}}, \ and\ \bibinfo {author} {\bibfnamefont {B.}~\bibnamefont
  {Schwabe}},\ }\href {\doibase 10.1103/PhysRevD.98.043509} {\bibfield
  {journal} {\bibinfo  {journal} {Phys. Rev.}\ }\textbf {\bibinfo {volume}
  {D98}},\ \bibinfo {pages} {043509} (\bibinfo {year} {2018})},\ \Eprint
  {http://arxiv.org/abs/1804.09647} {arXiv:1804.09647 [astro-ph.CO]}
  \BibitemShut {NoStop}%
\bibitem [{\citenamefont {Mocz}\ \emph {et~al.}(2017)\citenamefont {Mocz},
  \citenamefont {Vogelsberger}, \citenamefont {Robles}, \citenamefont {Zavala},
  \citenamefont {Boylan-Kolchin}, \citenamefont {Fialkov},\ and\ \citenamefont
  {Hernquist}}]{Mocz2017}%
  \BibitemOpen
  \bibfield  {author} {\bibinfo {author} {\bibfnamefont {P.}~\bibnamefont
  {Mocz}}, \bibinfo {author} {\bibfnamefont {M.}~\bibnamefont {Vogelsberger}},
  \bibinfo {author} {\bibfnamefont {V.~H.}\ \bibnamefont {Robles}}, \bibinfo
  {author} {\bibfnamefont {J.}~\bibnamefont {Zavala}}, \bibinfo {author}
  {\bibfnamefont {M.}~\bibnamefont {Boylan-Kolchin}}, \bibinfo {author}
  {\bibfnamefont {A.}~\bibnamefont {Fialkov}}, \ and\ \bibinfo {author}
  {\bibfnamefont {L.}~\bibnamefont {Hernquist}},\ }\href {\doibase
  10.1093/mnras/stx1887} {\bibfield  {journal} {\bibinfo  {journal} {Monthly
  Notices of the Royal Astronomical Society}\ }\textbf {\bibinfo {volume}
  {471}},\ \bibinfo {pages} {4559} (\bibinfo {year} {2017})},\ \Eprint
  {http://arxiv.org/abs/https://academic.oup.com/mnras/article-pdf/471/4/4559/19609125/stx1887.pdf}
  {https://academic.oup.com/mnras/article-pdf/471/4/4559/19609125/stx1887.pdf}
  \BibitemShut {NoStop}%
\bibitem [{\citenamefont {Mocz}\ \emph {et~al.}(2018)\citenamefont {Mocz},
  \citenamefont {Lancaster}, \citenamefont {Fialkov}, \citenamefont {Becerra},\
  and\ \citenamefont {Chavanis}}]{Mocz2018}%
  \BibitemOpen
  \bibfield  {author} {\bibinfo {author} {\bibfnamefont {P.}~\bibnamefont
  {Mocz}}, \bibinfo {author} {\bibfnamefont {L.}~\bibnamefont {Lancaster}},
  \bibinfo {author} {\bibfnamefont {A.}~\bibnamefont {Fialkov}}, \bibinfo
  {author} {\bibfnamefont {F.}~\bibnamefont {Becerra}}, \ and\ \bibinfo
  {author} {\bibfnamefont {P.-H.}\ \bibnamefont {Chavanis}},\ }\href {\doibase
  10.1103/PhysRevD.97.083519} {\bibfield  {journal} {\bibinfo  {journal} {Phys.
  Rev. D}\ }\textbf {\bibinfo {volume} {97}},\ \bibinfo {pages} {083519}
  (\bibinfo {year} {2018})}\BibitemShut {NoStop}%
\bibitem [{\citenamefont {{Tkachev}}(1986)}]{Tkachev1986}%
  \BibitemOpen
  \bibfield  {author} {\bibinfo {author} {\bibfnamefont {I.~I.}\ \bibnamefont
  {{Tkachev}}},\ }\href@noop {} {\bibfield  {journal} {\bibinfo  {journal}
  {Soviet Astronomy Letters}\ }\textbf {\bibinfo {volume} {12}},\ \bibinfo
  {pages} {305} (\bibinfo {year} {1986})}\BibitemShut {NoStop}%
\bibitem [{\citenamefont {Tkachev}(1991)}]{Tkachev1991}%
  \BibitemOpen
  \bibfield  {author} {\bibinfo {author} {\bibfnamefont {I.}~\bibnamefont
  {Tkachev}},\ }\href {\doibase https://doi.org/10.1016/0370-2693(91)90330-S}
  {\bibfield  {journal} {\bibinfo  {journal} {Physics Letters B}\ }\textbf
  {\bibinfo {volume} {261}},\ \bibinfo {pages} {289} (\bibinfo {year}
  {1991})}\BibitemShut {NoStop}%
\bibitem [{\citenamefont {Levkov}\ \emph {et~al.}(2018)\citenamefont {Levkov},
  \citenamefont {Panin},\ and\ \citenamefont {Tkachev}}]{Levkov2018}%
  \BibitemOpen
  \bibfield  {author} {\bibinfo {author} {\bibfnamefont {D.~G.}\ \bibnamefont
  {Levkov}}, \bibinfo {author} {\bibfnamefont {A.~G.}\ \bibnamefont {Panin}}, \
  and\ \bibinfo {author} {\bibfnamefont {I.~I.}\ \bibnamefont {Tkachev}},\
  }\href {\doibase 10.1103/PhysRevLett.121.151301} {\bibfield  {journal}
  {\bibinfo  {journal} {Phys. Rev. Lett.}\ }\textbf {\bibinfo {volume} {121}},\
  \bibinfo {pages} {151301} (\bibinfo {year} {2018})}\BibitemShut {NoStop}%
\bibitem [{\citenamefont {{Eggemeier}}\ and\ \citenamefont
  {{Niemeyer}}(2019)}]{Eggemeier2019PRD}%
  \BibitemOpen
  \bibfield  {author} {\bibinfo {author} {\bibfnamefont {B.}~\bibnamefont
  {{Eggemeier}}}\ and\ \bibinfo {author} {\bibfnamefont {J.~C.}\ \bibnamefont
  {{Niemeyer}}},\ }\href {\doibase 10.1103/PhysRevD.100.063528} {\bibfield
  {journal} {\bibinfo  {journal} {\prd}\ }\textbf {\bibinfo {volume} {100}},\
  \bibinfo {eid} {063528} (\bibinfo {year} {2019})},\ \Eprint
  {http://arxiv.org/abs/1906.01348} {arXiv:1906.01348 [astro-ph.CO]}
  \BibitemShut {NoStop}%
\bibitem [{\citenamefont {{Press}}\ and\ \citenamefont
  {{Schechter}}(1974)}]{Press1974}%
  \BibitemOpen
  \bibfield  {author} {\bibinfo {author} {\bibfnamefont {W.~H.}\ \bibnamefont
  {{Press}}}\ and\ \bibinfo {author} {\bibfnamefont {P.}~\bibnamefont
  {{Schechter}}},\ }\href {\doibase 10.1086/152650} {\bibfield  {journal}
  {\bibinfo  {journal} {\apj}\ }\textbf {\bibinfo {volume} {187}},\ \bibinfo
  {pages} {425} (\bibinfo {year} {1974})}\BibitemShut {NoStop}%
\bibitem [{\citenamefont {Akrami}\ \emph {et~al.}(2020)\citenamefont {Akrami}
  \emph {et~al.}}]{Planck2018_inflation}%
  \BibitemOpen
  \bibfield  {author} {\bibinfo {author} {\bibfnamefont {Y.}~\bibnamefont
  {Akrami}} \emph {et~al.} (\bibinfo {collaboration} {Planck}),\ }\href
  {\doibase 10.1051/0004-6361/201833887} {\bibfield  {journal} {\bibinfo
  {journal} {Astron. Astrophys.}\ }\textbf {\bibinfo {volume} {641}},\ \bibinfo
  {pages} {A10} (\bibinfo {year} {2020})},\ \Eprint
  {http://arxiv.org/abs/1807.06211} {arXiv:1807.06211 [astro-ph.CO]}
  \BibitemShut {NoStop}%
\bibitem [{\citenamefont {Albrecht}\ \emph {et~al.}(1982)\citenamefont
  {Albrecht}, \citenamefont {Steinhardt}, \citenamefont {Turner},\ and\
  \citenamefont {Wilczek}}]{Albrecht:1982mp}%
  \BibitemOpen
  \bibfield  {author} {\bibinfo {author} {\bibfnamefont {A.}~\bibnamefont
  {Albrecht}}, \bibinfo {author} {\bibfnamefont {P.~J.}\ \bibnamefont
  {Steinhardt}}, \bibinfo {author} {\bibfnamefont {M.~S.}\ \bibnamefont
  {Turner}}, \ and\ \bibinfo {author} {\bibfnamefont {F.}~\bibnamefont
  {Wilczek}},\ }\href {\doibase 10.1103/PhysRevLett.48.1437} {\bibfield
  {journal} {\bibinfo  {journal} {Phys. Rev. Lett.}\ }\textbf {\bibinfo
  {volume} {48}},\ \bibinfo {pages} {1437} (\bibinfo {year}
  {1982})}\BibitemShut {NoStop}%
\bibitem [{\citenamefont {Ruffini}\ and\ \citenamefont
  {Bonazzola}(1969)}]{Ruffini1969}%
  \BibitemOpen
  \bibfield  {author} {\bibinfo {author} {\bibfnamefont {R.}~\bibnamefont
  {Ruffini}}\ and\ \bibinfo {author} {\bibfnamefont {S.}~\bibnamefont
  {Bonazzola}},\ }\href {\doibase 10.1103/PhysRev.187.1767} {\bibfield
  {journal} {\bibinfo  {journal} {Phys. Rev.}\ }\textbf {\bibinfo {volume}
  {187}},\ \bibinfo {pages} {1767} (\bibinfo {year} {1969})}\BibitemShut
  {NoStop}%
\bibitem [{\citenamefont {Nambu}\ and\ \citenamefont
  {Sasaki}(1990)}]{Nambu1990QuantumPerturbations}%
  \BibitemOpen
  \bibfield  {author} {\bibinfo {author} {\bibfnamefont {Y.}~\bibnamefont
  {Nambu}}\ and\ \bibinfo {author} {\bibfnamefont {M.}~\bibnamefont {Sasaki}},\
  }\href {\doibase 10.1103/PhysRevD.42.3918} {\bibfield  {journal} {\bibinfo
  {journal} {Physical Review D}\ }\textbf {\bibinfo {volume} {42}},\ \bibinfo
  {pages} {3918} (\bibinfo {year} {1990})}\BibitemShut {NoStop}%
\bibitem [{\citenamefont {Schwabe}\ \emph {et~al.}(2020)\citenamefont
  {Schwabe}, \citenamefont {Gosenca}, \citenamefont {Behrens}, \citenamefont
  {Niemeyer},\ and\ \citenamefont {Easther}}]{Schwabe2020}%
  \BibitemOpen
  \bibfield  {author} {\bibinfo {author} {\bibfnamefont {B.}~\bibnamefont
  {Schwabe}}, \bibinfo {author} {\bibfnamefont {M.}~\bibnamefont {Gosenca}},
  \bibinfo {author} {\bibfnamefont {C.}~\bibnamefont {Behrens}}, \bibinfo
  {author} {\bibfnamefont {J.~C.}\ \bibnamefont {Niemeyer}}, \ and\ \bibinfo
  {author} {\bibfnamefont {R.}~\bibnamefont {Easther}},\ }\href {\doibase
  10.1103/PhysRevD.102.083518} {\bibfield  {journal} {\bibinfo  {journal}
  {Phys. Rev. D}\ }\textbf {\bibinfo {volume} {102}},\ \bibinfo {pages}
  {083518} (\bibinfo {year} {2020})}\BibitemShut {NoStop}%
\bibitem [{\citenamefont {Guzman}\ and\ \citenamefont
  {Urena-Lopez}(2004)}]{Guzman2004}%
  \BibitemOpen
  \bibfield  {author} {\bibinfo {author} {\bibfnamefont {F.~S.}\ \bibnamefont
  {Guzman}}\ and\ \bibinfo {author} {\bibfnamefont {L.~A.}\ \bibnamefont
  {Urena-Lopez}},\ }\href {\doibase 10.1103/PhysRevD.69.124033} {\bibfield
  {journal} {\bibinfo  {journal} {Phys. Rev. D}\ }\textbf {\bibinfo {volume}
  {D69}},\ \bibinfo {pages} {124033} (\bibinfo {year} {2004})}\BibitemShut
  {NoStop}%
\bibitem [{\citenamefont {Hahn}\ and\ \citenamefont {Abel}(2011)}]{Hahn2011}%
  \BibitemOpen
  \bibfield  {author} {\bibinfo {author} {\bibfnamefont {O.}~\bibnamefont
  {Hahn}}\ and\ \bibinfo {author} {\bibfnamefont {T.}~\bibnamefont {Abel}},\
  }\href {\doibase 10.1111/j.1365-2966.2011.18820.x} {\bibfield  {journal}
  {\bibinfo  {journal} {Monthly Notices of the Royal Astronomical Society}\
  }\textbf {\bibinfo {volume} {415}},\ \bibinfo {pages} {2101} (\bibinfo {year}
  {2011})}\BibitemShut {NoStop}%
\bibitem [{\citenamefont {Hui}\ \emph {et~al.}(2017)\citenamefont {Hui},
  \citenamefont {Ostriker}, \citenamefont {Tremaine},\ and\ \citenamefont
  {Witten}}]{Hui2017}%
  \BibitemOpen
  \bibfield  {author} {\bibinfo {author} {\bibfnamefont {L.}~\bibnamefont
  {Hui}}, \bibinfo {author} {\bibfnamefont {J.~P.}\ \bibnamefont {Ostriker}},
  \bibinfo {author} {\bibfnamefont {S.}~\bibnamefont {Tremaine}}, \ and\
  \bibinfo {author} {\bibfnamefont {E.}~\bibnamefont {Witten}},\ }\href
  {\doibase 10.1103/PhysRevD.95.043541} {\bibfield  {journal} {\bibinfo
  {journal} {Phys. Rev. D}\ }\textbf {\bibinfo {volume} {95}},\ \bibinfo
  {pages} {043541} (\bibinfo {year} {2017})}\BibitemShut {NoStop}%
\bibitem [{\citenamefont {Chen}\ \emph {et~al.}(2020)\citenamefont {Chen},
  \citenamefont {Du}, \citenamefont {Lentz}, \citenamefont {Marsh},\ and\
  \citenamefont {Niemeyer}}]{Jiajun2020}%
  \BibitemOpen
  \bibfield  {author} {\bibinfo {author} {\bibfnamefont {J.}~\bibnamefont
  {Chen}}, \bibinfo {author} {\bibfnamefont {X.}~\bibnamefont {Du}}, \bibinfo
  {author} {\bibfnamefont {E.~W.}\ \bibnamefont {Lentz}}, \bibinfo {author}
  {\bibfnamefont {D.~J.~E.}\ \bibnamefont {Marsh}}, \ and\ \bibinfo {author}
  {\bibfnamefont {J.~C.}\ \bibnamefont {Niemeyer}},\ }\href@noop {} {\
  (\bibinfo {year} {2020})},\ \Eprint {http://arxiv.org/abs/2011.01333}
  {arXiv:2011.01333 [astro-ph.CO]} \BibitemShut {NoStop}%
\bibitem [{\citenamefont {Schneider}\ \emph {et~al.}(2013)\citenamefont
  {Schneider}, \citenamefont {Smith},\ and\ \citenamefont
  {Reed}}]{Schneider2013}%
  \BibitemOpen
  \bibfield  {author} {\bibinfo {author} {\bibfnamefont {A.}~\bibnamefont
  {Schneider}}, \bibinfo {author} {\bibfnamefont {R.~E.}\ \bibnamefont
  {Smith}}, \ and\ \bibinfo {author} {\bibfnamefont {D.}~\bibnamefont {Reed}},\
  }\href {\doibase 10.1093/mnras/stt829} {\bibfield  {journal} {\bibinfo
  {journal} {Monthly Notices of the Royal Astronomical Society}\ }\textbf
  {\bibinfo {volume} {433}},\ \bibinfo {pages} {1573} (\bibinfo {year}
  {2013})}\BibitemShut {NoStop}%
\bibitem [{\citenamefont {Schneider}(2015)}]{Schneider2015}%
  \BibitemOpen
  \bibfield  {author} {\bibinfo {author} {\bibfnamefont {A.}~\bibnamefont
  {Schneider}},\ }\href {\doibase 10.1093/mnras/stv1169} {\bibfield  {journal}
  {\bibinfo  {journal} {Monthly Notices of the Royal Astronomical Society}\
  }\textbf {\bibinfo {volume} {451}},\ \bibinfo {pages} {3117} (\bibinfo {year}
  {2015})}\BibitemShut {NoStop}%
\bibitem [{\citenamefont {Liebling}\ and\ \citenamefont
  {Palenzuela}(2012)}]{Liebling2012}%
  \BibitemOpen
  \bibfield  {author} {\bibinfo {author} {\bibfnamefont {S.~L.}\ \bibnamefont
  {Liebling}}\ and\ \bibinfo {author} {\bibfnamefont {C.}~\bibnamefont
  {Palenzuela}},\ }\href {\doibase 10.12942/lrr-2012-6} {\bibfield  {journal}
  {\bibinfo  {journal} {Living Rev. Rel.}\ }\textbf {\bibinfo {volume} {15}},\
  \bibinfo {pages} {6} (\bibinfo {year} {2012})},\ \Eprint
  {http://arxiv.org/abs/1202.5809} {arXiv:1202.5809 [gr-qc]} \BibitemShut
  {NoStop}%
\bibitem [{\citenamefont {Seidel}\ and\ \citenamefont
  {Suen}(1990)}]{Seidel1990}%
  \BibitemOpen
  \bibfield  {author} {\bibinfo {author} {\bibfnamefont {E.}~\bibnamefont
  {Seidel}}\ and\ \bibinfo {author} {\bibfnamefont {W.-M.}\ \bibnamefont
  {Suen}},\ }\href {\doibase 10.1103/PhysRevD.42.384} {\bibfield  {journal}
  {\bibinfo  {journal} {Phys. Rev. D}\ }\textbf {\bibinfo {volume} {42}},\
  \bibinfo {pages} {384} (\bibinfo {year} {1990})}\BibitemShut {NoStop}%
\bibitem [{\citenamefont {Padilla}\ \emph {et~al.}(2021)\citenamefont
  {Padilla}, \citenamefont {Hidalgo},\ and\ \citenamefont
  {Malik}}]{Padilla:2021zgm}%
  \BibitemOpen
  \bibfield  {author} {\bibinfo {author} {\bibfnamefont {L.~E.}\ \bibnamefont
  {Padilla}}, \bibinfo {author} {\bibfnamefont {J.~C.}\ \bibnamefont
  {Hidalgo}}, \ and\ \bibinfo {author} {\bibfnamefont {K.~A.}\ \bibnamefont
  {Malik}},\ }\href@noop {} {\  (\bibinfo {year} {2021})},\ \Eprint
  {http://arxiv.org/abs/2110.14584} {arXiv:2110.14584 [astro-ph.CO]}
  \BibitemShut {NoStop}%
\bibitem [{\citenamefont {{Helfer}}\ \emph {et~al.}(2017)\citenamefont
  {{Helfer}}, \citenamefont {{Marsh}}, \citenamefont {{Clough}}, \citenamefont
  {{Fairbairn}}, \citenamefont {{Lim}},\ and\ \citenamefont
  {{Becerril}}}]{Helfer2017}%
  \BibitemOpen
  \bibfield  {author} {\bibinfo {author} {\bibfnamefont {T.}~\bibnamefont
  {{Helfer}}}, \bibinfo {author} {\bibfnamefont {D.~J.~E.}\ \bibnamefont
  {{Marsh}}}, \bibinfo {author} {\bibfnamefont {K.}~\bibnamefont {{Clough}}},
  \bibinfo {author} {\bibfnamefont {M.}~\bibnamefont {{Fairbairn}}}, \bibinfo
  {author} {\bibfnamefont {E.~A.}\ \bibnamefont {{Lim}}}, \ and\ \bibinfo
  {author} {\bibfnamefont {R.}~\bibnamefont {{Becerril}}},\ }\href {\doibase
  10.1088/1475-7516/2017/03/055} {\bibfield  {journal} {\bibinfo  {journal}
  {\jcap}\ }\textbf {\bibinfo {volume} {2017}},\ \bibinfo {eid} {055} (\bibinfo
  {year} {2017})},\ \Eprint {http://arxiv.org/abs/1609.04724} {arXiv:1609.04724
  [astro-ph.CO]} \BibitemShut {NoStop}%
\bibitem [{\citenamefont {Hooper}\ \emph {et~al.}(2019)\citenamefont {Hooper},
  \citenamefont {Krnjaic},\ and\ \citenamefont {McDermott}}]{Hooper2019}%
  \BibitemOpen
  \bibfield  {author} {\bibinfo {author} {\bibfnamefont {D.}~\bibnamefont
  {Hooper}}, \bibinfo {author} {\bibfnamefont {G.}~\bibnamefont {Krnjaic}}, \
  and\ \bibinfo {author} {\bibfnamefont {S.~D.}\ \bibnamefont {McDermott}},\
  }\href {\doibase 10.1007/JHEP08(2019)001} {\bibfield  {journal} {\bibinfo
  {journal} {JHEP}\ }\textbf {\bibinfo {volume} {08}},\ \bibinfo {pages} {001}
  (\bibinfo {year} {2019})},\ \Eprint {http://arxiv.org/abs/1905.01301}
  {arXiv:1905.01301 [hep-ph]} \BibitemShut {NoStop}%
\bibitem [{\citenamefont {Carr}\ and\ \citenamefont
  {Hawking}(1974)}]{Carr1974}%
  \BibitemOpen
  \bibfield  {author} {\bibinfo {author} {\bibfnamefont {B.~J.}\ \bibnamefont
  {Carr}}\ and\ \bibinfo {author} {\bibfnamefont {S.~W.}\ \bibnamefont
  {Hawking}},\ }\href {\doibase 10.1093/mnras/168.2.399} {\bibfield  {journal}
  {\bibinfo  {journal} {Monthly Notices of the Royal Astronomical Society}\
  }\textbf {\bibinfo {volume} {168}},\ \bibinfo {pages} {399} (\bibinfo {year}
  {1974})},\ \Eprint
  {http://arxiv.org/abs/https://academic.oup.com/mnras/article-pdf/168/2/399/8079885/mnras168-0399.pdf}
  {https://academic.oup.com/mnras/article-pdf/168/2/399/8079885/mnras168-0399.pdf}
  \BibitemShut {NoStop}%
\bibitem [{\citenamefont {{Carr}}(1975)}]{Carr1975}%
  \BibitemOpen
  \bibfield  {author} {\bibinfo {author} {\bibfnamefont {B.~J.}\ \bibnamefont
  {{Carr}}},\ }\href {\doibase 10.1086/153853} {\bibfield  {journal} {\bibinfo
  {journal} {\apj}\ }\textbf {\bibinfo {volume} {201}},\ \bibinfo {pages} {1}
  (\bibinfo {year} {1975})}\BibitemShut {NoStop}%
\bibitem [{\citenamefont {{Green}}\ \emph {et~al.}(1997)\citenamefont
  {{Green}}, \citenamefont {{Liddle}},\ and\ \citenamefont
  {{Riotto}}}]{Green1997}%
  \BibitemOpen
  \bibfield  {author} {\bibinfo {author} {\bibfnamefont {A.~M.}\ \bibnamefont
  {{Green}}}, \bibinfo {author} {\bibfnamefont {A.~R.}\ \bibnamefont
  {{Liddle}}}, \ and\ \bibinfo {author} {\bibfnamefont {A.}~\bibnamefont
  {{Riotto}}},\ }\href {\doibase 10.1103/PhysRevD.56.7559} {\bibfield
  {journal} {\bibinfo  {journal} {\prd}\ }\textbf {\bibinfo {volume} {56}},\
  \bibinfo {pages} {7559} (\bibinfo {year} {1997})},\ \Eprint
  {http://arxiv.org/abs/astro-ph/9705166} {arXiv:astro-ph/9705166 [astro-ph]}
  \BibitemShut {NoStop}%
\bibitem [{\citenamefont {{Green}}\ and\ \citenamefont
  {{Malik}}(2001)}]{Green2001}%
  \BibitemOpen
  \bibfield  {author} {\bibinfo {author} {\bibfnamefont {A.~M.}\ \bibnamefont
  {{Green}}}\ and\ \bibinfo {author} {\bibfnamefont {K.~A.}\ \bibnamefont
  {{Malik}}},\ }\href {\doibase 10.1103/PhysRevD.64.021301} {\bibfield
  {journal} {\bibinfo  {journal} {\prd}\ }\textbf {\bibinfo {volume} {64}},\
  \bibinfo {eid} {021301} (\bibinfo {year} {2001})},\ \Eprint
  {http://arxiv.org/abs/hep-ph/0008113} {arXiv:hep-ph/0008113 [hep-ph]}
  \BibitemShut {NoStop}%
\bibitem [{\citenamefont {Hawking}(1975)}]{Hawking1975}%
  \BibitemOpen
  \bibfield  {author} {\bibinfo {author} {\bibfnamefont {S.~W.}\ \bibnamefont
  {Hawking}},\ }\href {\doibase 10.1007/BF02345020} {\bibfield  {journal}
  {\bibinfo  {journal} {Commun. Math. Phys.}\ }\textbf {\bibinfo {volume}
  {43}},\ \bibinfo {pages} {199} (\bibinfo {year} {1975})},\ \bibinfo {note}
  {[Erratum: Commun.Math.Phys. 46, 206 (1976)]}\BibitemShut {NoStop}%
\bibitem [{\citenamefont {Fujita}\ \emph {et~al.}(2014)\citenamefont {Fujita},
  \citenamefont {Harigaya}, \citenamefont {Kawasaki},\ and\ \citenamefont
  {Matsuda}}]{Fujita2014}%
  \BibitemOpen
  \bibfield  {author} {\bibinfo {author} {\bibfnamefont {T.}~\bibnamefont
  {Fujita}}, \bibinfo {author} {\bibfnamefont {K.}~\bibnamefont {Harigaya}},
  \bibinfo {author} {\bibfnamefont {M.}~\bibnamefont {Kawasaki}}, \ and\
  \bibinfo {author} {\bibfnamefont {R.}~\bibnamefont {Matsuda}},\ }\href
  {\doibase 10.1103/PhysRevD.89.103501} {\bibfield  {journal} {\bibinfo
  {journal} {Phys. Rev. D}\ }\textbf {\bibinfo {volume} {89}},\ \bibinfo
  {pages} {103501} (\bibinfo {year} {2014})}\BibitemShut {NoStop}%
\bibitem [{\citenamefont {Lennon}\ \emph {et~al.}(2018)\citenamefont {Lennon},
  \citenamefont {March-Russell}, \citenamefont {Petrossian-Byrne},\ and\
  \citenamefont {Tillim}}]{Lennon2018}%
  \BibitemOpen
  \bibfield  {author} {\bibinfo {author} {\bibfnamefont {O.}~\bibnamefont
  {Lennon}}, \bibinfo {author} {\bibfnamefont {J.}~\bibnamefont
  {March-Russell}}, \bibinfo {author} {\bibfnamefont {R.}~\bibnamefont
  {Petrossian-Byrne}}, \ and\ \bibinfo {author} {\bibfnamefont
  {H.}~\bibnamefont {Tillim}},\ }\href {\doibase 10.1088/1475-7516/2018/04/009}
  {\bibfield  {journal} {\bibinfo  {journal} {Journal of Cosmology and
  Astroparticle Physics}\ }\textbf {\bibinfo {volume} {2018}},\ \bibinfo
  {pages} {009} (\bibinfo {year} {2018})}\BibitemShut {NoStop}%
\bibitem [{\citenamefont {Allahverdi}\ \emph {et~al.}(2018)\citenamefont
  {Allahverdi}, \citenamefont {Dent},\ and\ \citenamefont
  {Osinski}}]{Allahverdi2018}%
  \BibitemOpen
  \bibfield  {author} {\bibinfo {author} {\bibfnamefont {R.}~\bibnamefont
  {Allahverdi}}, \bibinfo {author} {\bibfnamefont {J.}~\bibnamefont {Dent}}, \
  and\ \bibinfo {author} {\bibfnamefont {J.}~\bibnamefont {Osinski}},\ }\href
  {\doibase 10.1103/PhysRevD.97.055013} {\bibfield  {journal} {\bibinfo
  {journal} {Phys. Rev. D}\ }\textbf {\bibinfo {volume} {97}},\ \bibinfo
  {pages} {055013} (\bibinfo {year} {2018})}\BibitemShut {NoStop}%
\bibitem [{\citenamefont {Morrison}\ \emph {et~al.}(2019)\citenamefont
  {Morrison}, \citenamefont {Profumo},\ and\ \citenamefont
  {Yu}}]{Morrison2019}%
  \BibitemOpen
  \bibfield  {author} {\bibinfo {author} {\bibfnamefont {L.}~\bibnamefont
  {Morrison}}, \bibinfo {author} {\bibfnamefont {S.}~\bibnamefont {Profumo}}, \
  and\ \bibinfo {author} {\bibfnamefont {Y.}~\bibnamefont {Yu}},\ }\href
  {\doibase 10.1088/1475-7516/2019/05/005} {\bibfield  {journal} {\bibinfo
  {journal} {Journal of Cosmology and Astroparticle Physics}\ }\textbf
  {\bibinfo {volume} {2019}},\ \bibinfo {pages} {005} (\bibinfo {year}
  {2019})}\BibitemShut {NoStop}%
\bibitem [{\citenamefont {Masina}(2020)}]{Masina2020}%
  \BibitemOpen
  \bibfield  {author} {\bibinfo {author} {\bibfnamefont {I.}~\bibnamefont
  {Masina}},\ }\href {\doibase 10.1140/epjp/s13360-020-00564-9} {\bibfield
  {journal} {\bibinfo  {journal} {Eur. Phys. J. Plus}\ }\textbf {\bibinfo
  {volume} {135}},\ \bibinfo {pages} {552} (\bibinfo {year} {2020})},\ \Eprint
  {http://arxiv.org/abs/2004.04740} {arXiv:2004.04740 [hep-ph]} \BibitemShut
  {NoStop}%
\bibitem [{\citenamefont {Cheek}\ \emph {et~al.}(2021)\citenamefont {Cheek},
  \citenamefont {Heurtier}, \citenamefont {Perez-Gonzalez},\ and\ \citenamefont
  {Turner}}]{Cheek2021}%
  \BibitemOpen
  \bibfield  {author} {\bibinfo {author} {\bibfnamefont {A.}~\bibnamefont
  {Cheek}}, \bibinfo {author} {\bibfnamefont {L.}~\bibnamefont {Heurtier}},
  \bibinfo {author} {\bibfnamefont {Y.~F.}\ \bibnamefont {Perez-Gonzalez}}, \
  and\ \bibinfo {author} {\bibfnamefont {J.}~\bibnamefont {Turner}},\
  }\href@noop {} {\  (\bibinfo {year} {2021})},\ \Eprint
  {http://arxiv.org/abs/2107.00013} {arXiv:2107.00013 [hep-ph]} \BibitemShut
  {NoStop}%
\bibitem [{\citenamefont {Khlopov}\ \emph {et~al.}(1985)\citenamefont
  {Khlopov}, \citenamefont {Malomed},\ and\ \citenamefont
  {Zeldovich}}]{Khlopov1985}%
  \BibitemOpen
  \bibfield  {author} {\bibinfo {author} {\bibfnamefont {M.~Y.}\ \bibnamefont
  {Khlopov}}, \bibinfo {author} {\bibfnamefont {B.~A.}\ \bibnamefont
  {Malomed}}, \ and\ \bibinfo {author} {\bibfnamefont {Y.~B.}\ \bibnamefont
  {Zeldovich}},\ }\href {\doibase 10.1093/mnras/215.4.575} {\bibfield
  {journal} {\bibinfo  {journal} {Monthly Notices of the Royal Astronomical
  Society}\ }\textbf {\bibinfo {volume} {215}},\ \bibinfo {pages} {575}
  (\bibinfo {year} {1985})},\ \Eprint
  {http://arxiv.org/abs/https://academic.oup.com/mnras/article-pdf/215/4/575/4082842/mnras215-0575.pdf}
  {https://academic.oup.com/mnras/article-pdf/215/4/575/4082842/mnras215-0575.pdf}
  \BibitemShut {NoStop}%
\bibitem [{\citenamefont {{Harada}}\ \emph {et~al.}(2016)\citenamefont
  {{Harada}}, \citenamefont {{Yoo}}, \citenamefont {{Kohri}}, \citenamefont
  {{Nakao}},\ and\ \citenamefont {{Jhingan}}}]{Harada2016}%
  \BibitemOpen
  \bibfield  {author} {\bibinfo {author} {\bibfnamefont {T.}~\bibnamefont
  {{Harada}}}, \bibinfo {author} {\bibfnamefont {C.-m.}\ \bibnamefont {{Yoo}}},
  \bibinfo {author} {\bibfnamefont {K.}~\bibnamefont {{Kohri}}}, \bibinfo
  {author} {\bibfnamefont {K.-i.}\ \bibnamefont {{Nakao}}}, \ and\ \bibinfo
  {author} {\bibfnamefont {S.}~\bibnamefont {{Jhingan}}},\ }\href {\doibase
  10.3847/1538-4357/833/1/61} {\bibfield  {journal} {\bibinfo  {journal}
  {\apj}\ }\textbf {\bibinfo {volume} {833}},\ \bibinfo {eid} {61} (\bibinfo
  {year} {2016})},\ \Eprint {http://arxiv.org/abs/1609.01588} {arXiv:1609.01588
  [astro-ph.CO]} \BibitemShut {NoStop}%
\bibitem [{\citenamefont {{Hidalgo}}\ \emph {et~al.}(2017)\citenamefont
  {{Hidalgo}}, \citenamefont {{De Santiago}}, \citenamefont {{German}},
  \citenamefont {{Barbosa-Cendejas}},\ and\ \citenamefont
  {{Ruiz-Luna}}}]{Hidalgo2017}%
  \BibitemOpen
  \bibfield  {author} {\bibinfo {author} {\bibfnamefont {J.~C.}\ \bibnamefont
  {{Hidalgo}}}, \bibinfo {author} {\bibfnamefont {J.}~\bibnamefont {{De
  Santiago}}}, \bibinfo {author} {\bibfnamefont {G.}~\bibnamefont {{German}}},
  \bibinfo {author} {\bibfnamefont {N.}~\bibnamefont {{Barbosa-Cendejas}}}, \
  and\ \bibinfo {author} {\bibfnamefont {W.}~\bibnamefont {{Ruiz-Luna}}},\
  }\href {\doibase 10.1103/PhysRevD.96.063504} {\bibfield  {journal} {\bibinfo
  {journal} {\prd}\ }\textbf {\bibinfo {volume} {96}},\ \bibinfo {eid} {063504}
  (\bibinfo {year} {2017})},\ \Eprint {http://arxiv.org/abs/1705.02308}
  {arXiv:1705.02308 [astro-ph.CO]} \BibitemShut {NoStop}%
\bibitem [{\citenamefont {{Carr}}\ \emph {et~al.}(2018)\citenamefont {{Carr}},
  \citenamefont {{Dimopoulos}}, \citenamefont {{Owen}},\ and\ \citenamefont
  {{Tenkanen}}}]{Carr2018}%
  \BibitemOpen
  \bibfield  {author} {\bibinfo {author} {\bibfnamefont {B.}~\bibnamefont
  {{Carr}}}, \bibinfo {author} {\bibfnamefont {K.}~\bibnamefont
  {{Dimopoulos}}}, \bibinfo {author} {\bibfnamefont {C.}~\bibnamefont
  {{Owen}}}, \ and\ \bibinfo {author} {\bibfnamefont {T.}~\bibnamefont
  {{Tenkanen}}},\ }\href {\doibase 10.1103/PhysRevD.97.123535} {\bibfield
  {journal} {\bibinfo  {journal} {\prd}\ }\textbf {\bibinfo {volume} {97}},\
  \bibinfo {eid} {123535} (\bibinfo {year} {2018})},\ \Eprint
  {http://arxiv.org/abs/1804.08639} {arXiv:1804.08639 [astro-ph.CO]}
  \BibitemShut {NoStop}%
\bibitem [{\citenamefont {{Martin}}\ \emph {et~al.}(2020)\citenamefont
  {{Martin}}, \citenamefont {{Papanikolaou}},\ and\ \citenamefont
  {{Vennin}}}]{Martin2020}%
  \BibitemOpen
  \bibfield  {author} {\bibinfo {author} {\bibfnamefont {J.}~\bibnamefont
  {{Martin}}}, \bibinfo {author} {\bibfnamefont {T.}~\bibnamefont
  {{Papanikolaou}}}, \ and\ \bibinfo {author} {\bibfnamefont {V.}~\bibnamefont
  {{Vennin}}},\ }\href {\doibase 10.1088/1475-7516/2020/01/024} {\bibfield
  {journal} {\bibinfo  {journal} {\jcap}\ }\textbf {\bibinfo {volume} {2020}},\
  \bibinfo {eid} {024} (\bibinfo {year} {2020})},\ \Eprint
  {http://arxiv.org/abs/1907.04236} {arXiv:1907.04236 [astro-ph.CO]}
  \BibitemShut {NoStop}%
\bibitem [{\citenamefont {Jedamzik}\ \emph
  {et~al.}(2010{\natexlab{b}})\citenamefont {Jedamzik}, \citenamefont
  {Lemoine},\ and\ \citenamefont {Martin}}]{Jedamzik2010_grav}%
  \BibitemOpen
  \bibfield  {author} {\bibinfo {author} {\bibfnamefont {K.}~\bibnamefont
  {Jedamzik}}, \bibinfo {author} {\bibfnamefont {M.}~\bibnamefont {Lemoine}}, \
  and\ \bibinfo {author} {\bibfnamefont {J.}~\bibnamefont {Martin}},\ }\href
  {\doibase 10.1088/1475-7516/2010/04/021} {\bibfield  {journal} {\bibinfo
  {journal} {Journal of Cosmology and Astroparticle Physics}\ }\textbf
  {\bibinfo {volume} {2010}},\ \bibinfo {pages} {021} (\bibinfo {year}
  {2010}{\natexlab{b}})}\BibitemShut {NoStop}%
\bibitem [{\citenamefont {Aggarwal}\ \emph {et~al.}(2020)\citenamefont
  {Aggarwal} \emph {et~al.}}]{Aggarwal2020}%
  \BibitemOpen
  \bibfield  {author} {\bibinfo {author} {\bibfnamefont {N.}~\bibnamefont
  {Aggarwal}} \emph {et~al.},\ }\href@noop {} {\  (\bibinfo {year} {2020})},\
  \Eprint {http://arxiv.org/abs/2011.12414} {arXiv:2011.12414 [gr-qc]}
  \BibitemShut {NoStop}%
\bibitem [{\citenamefont {Hertzberg}\ and\ \citenamefont
  {Schiappacasse}(2018)}]{Hertzberg2018}%
  \BibitemOpen
  \bibfield  {author} {\bibinfo {author} {\bibfnamefont {M.~P.}\ \bibnamefont
  {Hertzberg}}\ and\ \bibinfo {author} {\bibfnamefont {E.~D.}\ \bibnamefont
  {Schiappacasse}},\ }\href {\doibase 10.1088/1475-7516/2018/11/004} {\bibfield
   {journal} {\bibinfo  {journal} {JCAP}\ }\textbf {\bibinfo {volume} {11}},\
  \bibinfo {pages} {004} (\bibinfo {year} {2018})},\ \Eprint
  {http://arxiv.org/abs/1805.00430} {arXiv:1805.00430 [hep-ph]} \BibitemShut
  {NoStop}%
\bibitem [{\citenamefont {Levkov}\ \emph {et~al.}(2020)\citenamefont {Levkov},
  \citenamefont {Panin},\ and\ \citenamefont {Tkachev}}]{Levkov2020}%
  \BibitemOpen
  \bibfield  {author} {\bibinfo {author} {\bibfnamefont {D.~G.}\ \bibnamefont
  {Levkov}}, \bibinfo {author} {\bibfnamefont {A.~G.}\ \bibnamefont {Panin}}, \
  and\ \bibinfo {author} {\bibfnamefont {I.~I.}\ \bibnamefont {Tkachev}},\
  }\href {\doibase 10.1103/PhysRevD.102.023501} {\bibfield  {journal} {\bibinfo
   {journal} {Phys. Rev. D}\ }\textbf {\bibinfo {volume} {102}},\ \bibinfo
  {pages} {023501} (\bibinfo {year} {2020})},\ \Eprint
  {http://arxiv.org/abs/2004.05179} {arXiv:2004.05179 [astro-ph.CO]}
  \BibitemShut {NoStop}%
\bibitem [{\citenamefont {{Turk}}\ \emph {et~al.}(2011)\citenamefont {{Turk}},
  \citenamefont {{Smith}}, \citenamefont {{Oishi}}, \citenamefont {{Skory}},
  \citenamefont {{Skillman}}, \citenamefont {{Abel}},\ and\ \citenamefont
  {{Norman}}}]{Turk2011}%
  \BibitemOpen
  \bibfield  {author} {\bibinfo {author} {\bibfnamefont {M.~J.}\ \bibnamefont
  {{Turk}}}, \bibinfo {author} {\bibfnamefont {B.~D.}\ \bibnamefont {{Smith}}},
  \bibinfo {author} {\bibfnamefont {J.~S.}\ \bibnamefont {{Oishi}}}, \bibinfo
  {author} {\bibfnamefont {S.}~\bibnamefont {{Skory}}}, \bibinfo {author}
  {\bibfnamefont {S.~W.}\ \bibnamefont {{Skillman}}}, \bibinfo {author}
  {\bibfnamefont {T.}~\bibnamefont {{Abel}}}, \ and\ \bibinfo {author}
  {\bibfnamefont {M.~L.}\ \bibnamefont {{Norman}}},\ }\href {\doibase
  10.1088/0067-0049/192/1/9} {\bibfield  {journal} {\bibinfo  {journal} {APJS}\
  }\textbf {\bibinfo {volume} {192}},\ \bibinfo {eid} {9} (\bibinfo {year}
  {2011})},\ \Eprint {http://arxiv.org/abs/1011.3514} {arXiv:1011.3514
  [astro-ph.IM]} \BibitemShut {NoStop}%
\end{thebibliography}%

\end{document}